# CHARACTERIZATION OF 1-ALKANOL + STRONGLY POLAR COMPOUND MIXTURES FROM THERMOPHYSICAL DATA AND THE APPLICATION OF THE KIRKWOOD-BUFF INTEGRALS AND KIRKWOOD-FRÖHLICH FORMALISMS.

JUAN ANTONIO GONZÁLEZ*, FERNANDO HEVIA, LUIS FELIPE SANZ, ISAÍAS GARCÍA DE LA FUENTE AND JOSÉ CARLOS COBOS

G.E.T.E.F., Departamento de Física Aplicada, Facultad de Ciencias, Universidad de Valladolid,  Paseo de Belén, 7, 47011 Valladolid, Spain,

*e-mail: jagl@termo.uva.es; Fax: +34-983-423136; Tel: +34-983-423757

**Abstract**

Mixtures formed by 1-alkanol and one strongly polar compound, nitromethane (NM), ethanenitrile (EtN), dimethyl sulfoxide (DMSO, sulfolane (SULF), nitrobenzene (NTBz) or benzonitrile (BzCN), have been investigated on the basis of a set of thermophysical data, which includes: excess molar functions, enthalpies, $H_m^E$, Gibbs energies, $G_m^E$, entropies, $TS_m^E$, isobaric heat capacities, $C_{pm}^E$, volumes, $V_m^E$; liquid-liquid equilibria (LLE), excess permittivies and deviations from the linearity of dynamic viscosities. In addition, calculations have been conducted to determine the Kirkwood-Buff integrals and the Kirkwood correlations factors, $g_K$, of the investigated mixtures. In the former case, DISQUAC has been employed for modeling the needed vapor-liquid equilibria data. Many systems under consideration are characterized by dipolar interactions between like molecules and have positive values of $H_m^E$, $C_{pm}^E$ and $TS_m^E$. On the other hand, alkanol-solvent interactions, for mixtures with a fixed 1-alkanol, become weakened in the sequence: DMSO $\approx$ SULF > EtN > NM > BzCN > NTBz. In systems with a given solvent, such interactions become also weaker when the chain length of the 1-alkanol is increased. Interestingly, the considered mixtures also show strong structural effects. Results on Kirkwood-Buff integrals reveal that nitriles are more preferred than nitroalkanes around a central alcohol molecule. Calculations on $g_K$ show that, in terms of the mixture polarization, the ssystems are rather unstructured, and that this trend becomes more important when the 1-alkanol size increases in solutions with a given solvent.

1.   **Introduction**

Nitromethane, (NM), ethanenitrile (EtN), dimethyl sulfoxide (DMSO), sulfolane (SULF), nitrobenzene (NTBz) or benzonitrile (BzCN) are polar compounds with very large dipole moments, $\mu$, (Table 1). For example, the $\mu$ value of sulfolane is 4.81 D [1]. These aprotic solvents have many applications. Sulfolane is useful in the oil industry for the recovery, by liquid extraction, of aromatic or saturated hydrocarbons [2,3]. Nitriles are used as starting materials in the synthesis of pesticides, fragances and pharmaceuticals [4]. NM is important in the manufacture of pesticides or drugs and NTBZ is essential for the aniline production. NM is also a high performance additive to fuel for internal combustion engines [5], particularly interesting since it is of relatively non-toxic nature. The considered solvents have a certain structure in liquid state, which can be ascribed to their high polarity. In fact, X-ray diffraction studies on DMSO point out that this pure liquid is structured due to dipole-dipole interactions [6]. Experiments using IR and Raman spectroscopic techniques indicate that liquid NM has molecules in monomeric state, and that self-associated dimers do not exist [7]. X-ray and neutron diffraction measurements and simulation calculations reveal that orientational correlations exist in pure NM resulting in antiparallel order of the neighboring molecules [8]. It is to be noted that sulfolane does not easily interact with others molecules due to the steric hindrance related to its globular shape which makes that only the negative end of its large $\mu$ is exposed [9-12].

In the past years, we have investigated the mixtures 1-alkanol + SULF [13], or + DMSO [14], or + EtN [15,16], or + NM [17], or + BzCN [18], or + NTBz [17] using DISQUAC [19,20] or ERAS [21], or the Flory model [22] or the concentration-concentration structure factor formalism, $S_{CC}(0)$, [23,24]. We extend now our studies by the application of the Kirkwood-Buff integrals (KBIs) formalism [25,26] and the Kirkwood-Fröhlich model [27-29] to the mentioned systems. The former theory takes into account fluctuations in the number of molecules of each component and the cross fluctuations, and the KBIs are derived from thermodynamic properties such vapour-liquid equilibria (VLE) and excess molar volumes, $V_m^E$. In the present work, VLE of the investigated systems were modelized by means of DISQUAC using mainly interaction parameters from the literature (see below). The Kirkwood-Fröhlich model allows calculate the Kirkwood's correlation factor, $g_K$, [27-29] an important magnitude which provides information about specific interactions in the liquid state.

The determination of KBIs using DISQUAC may be supported by the following considerations. (i) For the studied systems, the VLE data available are scarce and are at different temperatures. The DISQUAC application allows compare results on KBIs at the same temperature (here, 298.15 K). This is important since KBIs are sensitive to temperature changes. (ii) In previous studies [13,14,16-18], we have shown that DISQUAC is useful to represent

$G_m^E$ (excess molar Gibbs energies, see Figures included in supplementary material) and any type of phase equilibria, vapour-liquid, liquid-liquid, or solid-liquid equilibria. Therefore, one can expect that the KBIs obtained are useful for the research of these systems. (iii) It is crucial to note that the mentioned investigation is not conducted only in terms of the KBIs. In fact, one of the aims of the work is to evaluate if the information obtained from the KBI formalism and from the Kirkwood-Fröhlich model is consistent with that provided from usual thermophysical properties such as excess molar functions: enthalpies, $H_m^E$, $G_m^E$, $V_m^E$, entropies, $TS_m^E (= H_m^E - G_m^E)$, excess permittivities, $\varepsilon_r^E$, or deviations from linearity of dynamic viscosities, $\Delta\eta$, At this end, Table 2 contains data on $H_m^E$, $G_m^E$, $V_m^E$, and $TS_m^E$ values at 298.15 K for the systems under study, and Table 3 lists the corresponding $\varepsilon_r^E$ and $\Delta\eta$ values. We remark that, due to the lack of experimental data, the $G_m^E$ values were usually obtained from DISQUAC.

## 2. Models

### 2.1 Kirkwood-Buff integrals

These magnitudes can be determined from thermodynamic data by means of the following expressions [26,30]:

$$G_{11} = RT\kappa_T + \frac{x_2 \overline{V}_{m2}^2}{x_1 V_m D} - \frac{V_m}{x_1} \quad (1)$$

$$G_{22} = RT\kappa_T + \frac{x_1 \overline{V}_{m1}^2}{x_2 V_m D} - \frac{V_m}{x_2} \quad (2)$$

$$G_{12} = G_{21} = RT\kappa_T - \frac{\overline{V}_{m1}\overline{V}_{m2}}{V_m D} \quad (3)$$

In equations (1-3), $x_i$ and $\overline{V}_{mi}$ are the mole fraction and the partial molar volume of component i, respectively (i = 1,2); $V_m$ is the molar volume of the solution at the working temperature, $T$, and $\kappa_T$, the isothermal compressibility factor of the mixture. $D$ is defined as:

$$D = 1 + \frac{x_1 x_2}{RT}\left(\frac{\partial^2 G_m^E}{\partial x_1^2}\right)_{P,T} \quad (4)$$

The $G_{ij}$ quantities allow estimate the so-called linear coefficients of preferential solvation:

$$\delta_{ij} = x_i G_{ij} - x_i \sum_k x_k G_{kj} \qquad (5)$$

These magnitudes provide information about changes in the local mole fractions of compound i around a central j molecule [31].

*2.2    Kirkwood-Fröhlich model*

We summarize some important hypotheses of the model [27-29]. (i) A molecule of a certain polar compound is represented by a dipole moment inside a spherical cavity. (ii) The effect of the induced polarization of the molecules is macroscopically treated. At this end, it is assumed that the dipole is rigid (it only rotates) and that the cavity is filled by a continuous medium of relative permittivity $\varepsilon_r^\infty$ (the permittivity value at a high frequency at which only the induced polarizability contributes). (iii) Long-range interactions are considered macroscopically by assuming that the outside of the cavity is a continuous dielectric of permittivity $\varepsilon_r$. (iv) Effects due to short-range interactions are taken into account by means of $g_K$, which provides information on the deviations from randomness of the orientation of a dipole with respect to its neighbours. For a mixture, $g_K$ can be determined, in the framework of a one-fluid model [27], from macroscopic physical properties according to the expression [27-29]:

$$g_K = \frac{9 k_B T V_m \varepsilon_0 (\varepsilon_r - \varepsilon_r^\infty)(2\varepsilon_r + \varepsilon_r^\infty)}{N_A \mu^2 \varepsilon_r (\varepsilon_r^\infty + 2)^2} \qquad (6)$$

The symbols have their usual meaning: $k_B$ is Boltzmann's constant; $N_A$, the Avogadro's number; and $\varepsilon_0$, the vacuum permittivity. For polar compounds, $\varepsilon_r^\infty$ is estimated from the relation $\varepsilon_r^\infty = 1.1 n_D^2$ [32,33]. The dipole moment of the solution, $\mu$, is estimated from the equation [27]:

$$\mu^2 = x_1 \mu_1^2 + x_2 \mu_2^2 \qquad (7)$$

where $\mu_i$ stands for the dipole moment of component i (=1,2).

**3.    Results**

*3.1    Kirkwood-Buff integrals*

The $\kappa_T$ values of the mixtures under study were calculated assuming that they behave ideally with respect to this property. That is, $\kappa_T = \varphi_1 \kappa_{T1} + \varphi_2 \kappa_{T2}$, where $\varphi_i$ stands for the volume fraction of the component i of the system (= $x_i V_{mi} / (x_1 V_{m1} + x_2 V_{m2})$; $V_{mi}$ is the molar volume of component i), and $\kappa_{Ti}$, its isothermal compressibility. This assumption has not influence on the final values of the Kirkwood-Buff integrals [34,35]. In addition, the data available in the literature for the studied mixtures on excess isentropic or isothermal compressibilities show that these magnitudes are small in absolute value [36,37]. Values of $\kappa_{Ti}$ for the polar compounds considered in this work are listed in Table 1. For 1-alkanols, the $\kappa_{Ti}$ values are the same as in reference [38]. Kirkwood-Buff integrals were determined at 298.15 K. References for most of the experimental $V_m^E$ data needed for calculations are given in Table 2. In absence of accurate values for this property at 298.15 K, $V_m^E$ values at 303.15 K were used for a few systems (Table 2). For the 1-octanol + EtN system, $V_m^E$ was taken from reference [39], and for 1-propanol or 1-butanol + SULF mixtures from reference [40]. Most of the interactions parameters needed for the application of DISQUAC to calculate $D$ values (equation 4) were taken from the literature (see below). Only a few of them were modified, as it is now indicated.

*3.1.1    1-alkanol + ethanenitrile, or + nitromethane, or +DMSO*

There are three contacts in these systems: OH/CH$_2$ [41,42]; CN/CH$_2$ [43] and OH/CN [15], in EtN mixtures;  OH/CH$_2$; NO$_2$/CH$_2$ [44] and OH/NO$_2$ [17] in NM solutions, and OH/CH$_2$; SO/CH$_2$  and OH/SO [14] in systems including DMSO.

*3.1.2    1-alkanol + sulfolane*

We have here six contacts: OH/CH$_2$; OH/c-CH$_2$ [45,46]; c-SO$_2$/CH$_2$; c-SO$_2$/c-CH$_2$; CH$_2$/c-CH$_2$ and OH/c-SO$_2$ [13]. It is remarkable that, in our original work [13], the interaction parameters for CH$_2$/c-CH$_2$ contacts were neglected in view of the low values of the excess functions of cyclopentane + *n*-alkane systems. The initial first DIS parameters for the OH/c-SO$_2$ contacts in methanol or ethanol are haven slightly changed for a better representation of the $G_m^E$ of these systems (see Figure S4, supplementary material). *C*

*3.1.3    1-alkanol + benzonitrile, or + nitrobenzene*

There are also 6 contacts in these solutions. In mixtures with benzonitrile, the contacts are: OH/CH$_2$; OH/C$_6$H$_5$ [42,47]; CN/CH$_2$ [18]; CN/C$_6$H$_5$ [18]; C$_6$H$_5$/CH$_2$ [48] and OH/CN [18]. In systems with NTBz, the CN/CH$_2$, CN/C$_6$H$_5$ and OH/CN contacts are replaced by the NO$_2$/CH$_2$; NO$_2$/C$_6$H$_5$ and OH/NO$_2$  contacts [17]. The first dispersive interchange coefficients for the contacts OH/CN in systems with methanol or ethanol, or for the contacts OH/NO$_2$ in mixtures with ethanol, 1-propanol or 1-butanol were modified (Table 4) taking into account the activity coefficients at infinite dilution of the 1-alkanols available in the literature [49].

Results on Kirkwood Buff integrals and on linear coefficients of preferential solvation are listed in Table 5 (see Figures 1-5).

*3.2    Calculations on Kirkwood's correlation factor*

Dipole moments of the polar solvents are listed in Table 1. For 1-alkanols were taken from reference [33]. Due to the lack of reliable experimental $n_\text{D}$ data, results for the mixtures were obtained from [50]:

$$n_\text{D}^\text{id} = \left[ \varphi_1 \left( n_\text{D1} \right)^2 + \varphi_2 \left( n_\text{D2} \right)^2 \right]^{1/2} \tag{8}$$

That is, the mixtures were considered as ideal with respect to $n_\text{D}$. Most of the values of this magnitude for pure compounds, $n_\text{Di}$, were taken from reference [1]. For 1-propanol and BzCN at 303.15 K, $n_\text{Di}$ values were obtained from [51] and [52], respectively. For 1-heptanol at 298.15 K, the value used is that given in reference [53]. Molar volumes and permittivities of the solutions were determined from $V_\text{m}^\text{E}$ and $\varepsilon_\text{r}^\text{E}$ values obtained from references listed in Tables 2 and 3, respectively. For the 1-propanol + BzCN system at 303.15 K, $V_\text{m}^\text{E}$ data were taken from [54] and for the 1-pentanol + NTBz mixture at 293.15 K, from [55]. Permittivitties of pure compounds were taken from the original papers were $\varepsilon_\text{r}$ values for the mixtures are reported (Table 3). Results on $g_\text{K}$ are listed in Table 3 and represented in Figures 6-7.

## 5.    Discussion

Below, except when indicated, we are referring to excess molar functions at equimolar composition and 298.15 K. The number of C atoms in 1-alkanol is represented by $n_\text{OH}$.

*5.1    Polar compound + alkane mixtures*

Figure 8 shows values of the UCST for NM, or EtN or NTBz, or BzCN + *n*-alkane mixtures. We note that the critical temperatures of the systems with NM or EtN are rather high, which reveals the existence of strong dipolar interactions in the mentioned mixtures. The same can be stated for SULF, or DMSO solutions. In fact, the mixtures SULF + heptane, and DMSO + cyclohexane show miscibility gaps at 429.4 K for $x_1 \in [0.0637, 0.9603]$ [56] and at 323.15 K for $x_1 \in [0.012, 0.9377]$ [57,58], respectively. The UCST/K of the NM or EtN + cyclohexane systems is (in the same order): 365.8 [59] and 347.6 [60]. In view of these results, one can conclude that interactions between like molecules become weaker, in systems with a fixed alkane, along the sequence: SULF > DMSO > NM > EtN > NTBz > BzCN. This can be confirmed by studying the relative changes in intermolecular forces of homomorphic compounds, which can be estimated from the $\Delta\Delta H_{vap}$ magnitude, defined as [61-63]:

$$\Delta\Delta H_{vap} = \Delta H_{vap} \text{ (compound with a given polar group, X)} -$$
$$\Delta H_{vap} \text{ (homomorphic hydrocarbon)} \quad (9)$$

In this equation $\Delta H_{vap}$ is the standard enthalpy of vaporization at 298.15 K. The relative variation of $\Delta\Delta H_{vap}$ (in kJ·mol$^{-1}$) is as follows: 39.1 (SULF) > 36.6 (DMSO) > 28.6 (NM) > 23.6 (EtN) > 18 (NTBz) ≈ 17.7 (BzCN) (Table 1) and is in good agreement with the observed relative change of the UCSTs.

The impact of polarity on bulk properties can be examined through the effective dipole moment, $\bar{\mu}$, defined by [64-67]:

$$\bar{\mu} = \left[ \frac{\mu^2 N_A}{4\pi\varepsilon_0 V_m k_B T} \right]^{1/2} \quad (10)$$

Values of $\bar{\mu}$ for the pure polar solvents are listed in Table 1. Such values suggest that dipolar interactions are very similar for NM, EtN, SULF or DMSO, and weaker for NTBz and BzCN. Calculations on the potential energy of dipole-dipole interactions in a pure polar liquid, roughly proportional to $(-\mu^4 / V_m^2)$ [68], confirm this point. It seems that, e.g. mixtures containing NM or EtN differ by dispersive interactions.

*5.2    1-alkanol + polar compound*

Experimental UCST results for 1-alkanol + NM, or + EtN, or + SULF mixtures are shown in the Figure 8. They deserve two comments. (i) The existence of miscibility gaps at temperatures not far from 298.15 K for systems containing an alcohol of medium size (e.g., UCST(1-hexanol + NM) = 308.75 K [69]) underlines that dipolar interactions between like molecules are still relevant. (ii) For a given polar compound, UCST(1-alkanol) < UCST(homomorphic *n*-alkane), which reveals that interactions between unlike molecules are more relevant in the alcoholic solutions.

*5.2.1    Kirkwood-Buff integrals and derived quantities*

Interestingly, for systems containing NM, or NTBz, or BzCN or SULF or for the mixture 1-octanol + EtN, the $G_{ii}$ (i =1,2) curves show a rather high maximum and the $G_{ij}$ curve a deep minimum (Figure 1). Particularly, for the 1-butanol + nitromethane mixture at 298.15 K, $G_{11}(x_1 = 0.37) = 2941$; $G_{22}(x_1 = 0.41) = 3577$ and $G_{12}(x_1 = 0.39) = -3244$ (all values in

cm$^3$·mol$^{-1}$). One can then conclude that the mentioned solutions are mainly characterized by interactions between like molecules. Results on $G_{ij}$ curves are available in the literature for 1-alkanol + heptane systems at 313.15 K [70]. Their concentration dependence is rather different to that described above since, in alkane systems, only alkanol-alkanol interactions are significant. Thus, for the 1-butanol system, the $G_{11}$ curve shows a maximum value of ≈ 8000 cm$^3$·mol$^{-1}$ at $x_1$ ≈ 0.10, while the minimum value of the $G_{12}$ curve, encountered at $x_1$ ≈ 0.15, is ≈ −1000 cm$^3$·mol$^{-1}$. In comparison, the $G_{22}$ values are more or less negligible [70]. According to the $G_{ij}$ values of 1-alkanol + polar solvent (≠DMSO) mixtures, results for $\delta_{21}$ are negative and for $\delta_{22}$ are positive (Table 5, Figures 2,3,5). In addition, the absolute values of these magnitudes are usually large, indicating that interactions between like molecules are dominant. This is consistent with our previous application of the $S_{CC}(0)$ formalism to 1-alkanol + 1-nitroalkane mixtures [17], which showed that these systems are characterized by homocoordination. Thus, for the ethanol + NM system at $x_1$ = 0.47, $S_{CC}(0)$ = 4.9 [17]. Inspection of Table 5 allows state some interesting conclusions. (i) For a given polar compound, $\delta_{21}$ decreases and $\delta_{22}$ increases when $n_{OH}$ is increased. That is, interactions between unlike molecules become less relevant at such condition. In terms of the $S_{CC}(0)$ magnitude, this means that homocoordination increases. Consequently, the $S_{CC}(0)$ value of the 1-propanol + NM mixture (7.14 at $x_1$ = 0.38 [17]) is larger than that of the ethanol solution. Since $|\delta_{21}|$ and $\delta_{22}$ values of mixtures containing NM, NTBz, BzCN, SULF, or of the 1-octanol + EtN system are rather large, and the linear coefficients of preferential solvation, $\delta_{ij}$, reflect changes in the local mole fractions of component i around a central molecule of type j, the local mole fractions will largely differ from the bulk ones. This is in agreement with the observed variation of UCST with the alkanol size for solutions with NM, EtN, or SULF (Figure 8). We remark that lower $G_m^E$ values are encountered for mixtures involving shorter 1-alkanols (Table 2), and that the corresponding values of the Gibbs energy of mixing ($G_m^M = G_m^E + G_m^{id}$) are then more negative. Compounds mix better. (ii) For a fixed 1-alkanol, the replacement of nitromethane or nitrobenzene by the corresponding nitrile (ethanenitrile or benzonitrile) leads to higher $\delta_{21}$ values and lower $\delta_{22}$ values. In other words, interactions between unlike molecules become relatively more important in nitrile systems. The Kirkwood-Buff theory provides $G_{ia} - G_{ib} > 0$ as a criterion to know if a solvent "a" is preferred over a solvent "b" in the vicinity of a given solute "i" [71,72]. Calculations using values of the Kirkwood-Buff integrals listed in Table 5 show that for a = NM, b = EtN or a = NTBz, b = BzCN, the differences $G_{ia} - G_{ib}$ are negative. That is, nitriles are more preferred than nitroalkanes around a given alkanol molecule.

Accordingly, $S_{CC}(0)$ values of 1-alkanol + EtN mixtures are lower than those of 1-alkanol + NM systems. For the sake of comparison, we provide the result for the ethanol + EtN mixture, $S_{CC}(0)(x_1 = 0.5) = 0.79$ [16]. Note that for a given 1-alkanol, UCST(NM) > UCST(EtN) (Figure 8). (iii) The same procedure shows that sulfolane molecules are more preferred around a central 1-alkanol molecule ($n_{OH} = 1$-4) than NM molecules, which is in agreement with the fact that, for $n_{OH} = 4$, UCST/K = 291.1 (NM) [58,73] > 285.6 (SULF) [13]. However, this trend may be the opposite for mixtures involving long chain 1-alkanols. (Figure 8). (iii) DMSO systems or mixtures formed by EtN and shorter 1-alkanols are characterized by low values of $|\delta_{21}|$ and $\delta_{22}$ (Table 5, Figure 4). Therefore, local mole fractions do not differ from the bulk ones. Mixtures containing 1-alkanols and secondary or tertiary amides [31,74-76], or 1-alkanols and pyridine [77], or the 1-propanol + tetrahydrofuran system [78,79] behave similarly. For example, in the case of the ethanol + *N,N*-dimethylformamide mixture at 313.15 K and $x_1 = 0.4$, $\delta_{21}$ and $\delta_{22}$ are, respectively, 6.5 and $-0.48$ cm$^3 \cdot$mol$^{-1}$ [31]. This can be interpreted assuming that, along the mixing process, a large number of interactions between like molecules are broken while, simultaneously, a large number of alkanol-solvent interactions are created. However, the low $|\delta_{21}|$ values indicate that the latter interactions do not involve a large number of unlike molecules. For example, NMR and permittivity measurements for the methanol + DMSO system suggest that only dimers or trimers exist in the solution [80]. Low $|\delta_{21}|$ values mean that the radius of the solvation microsphere is small and the distribution of the molecules in the solution is nearly random. In such cases, the mixture structure can be ascribed to the existence of orientational effects [74]. In order to explore the relevance of these effects in the present systems, we have shortly applied the Flory model [22] to some 1-alkanol + DMSO mixtures. Below, we provide the interaction parameter, $X_{12}$, and corresponding standard relative deviations, $\sigma_r(H_m^E)$, for $H_m^E$, in order to characterize the differences between experimental and theoretical results for this magnitude. The $\sigma_r(H_m^E)$ values are calculated from:

$$\sigma_r(H_m^E) = \{\frac{1}{N}\sum \left[\frac{H_{m,exp}^E - H_{m,calc}^E}{H_{m,exp}^E}\right]^2\}^{1/2} \quad (11)$$

where $N$ stands for the number of data points. More details on this type of calculations can be found elsewhere [16]. Using $H_m^E$ data from [81], we have obtained $\sigma_r(H_m^E)$ (DMSO) = 0.174 ($n_{OH} = 4$, $X_{12} = 54.83$ J·cm$^{-3}$); 0.085 ($n_{OH} = 6$, $X_{12} = 62$ J·cm$^{-3}$); 0.085 ($n_{OH} = 8$, $X_{12} = 60.65$ J·cm$^{-3}$); 0.074 ($n_{OH} = 10$, $X_{12} = 58.22$ J·cm$^{-3}$). These results indicate that orientational effects are rather weak in systems with, say, $n_{OH} \geq 5$. Stronger orientational effects are encountered in solutions with shorter 1-alkanols. Regarding 1-alkanol + EtN mixtures, our previous study [16] using the Flory model reveals that orientational effects are relevant in the methanol system or in those solutions at temperatures close to the UCST ($\sigma_r(H_m^E) = 0.169$; $n_{OH} = 10$). For the remainder systems, the orientational effects are weak ($\sigma_r(H_m^E) = 0.077$ ($n_{OH} = 2$); 0.094 ($n_{OH} = 4$)) [16].

### 5.2.2 *Enthalpy of alkanol-solvent interactions*

In some previous works, we have evaluated the enthalpy of the interactions between 1-alkanols and polar compounds including a functional group X (= $NO_2$; CN; O) [16,17,82] (termed as $\Delta H_{OH-X}$). Here, we follow the same procedure to determine $\Delta H_{OH-X}$ for X = SO and c-$SO_2$ (Table 6) and for ethanenitrile systems with $n_{OH} = 5,6$ (values not previously reported). Briefly, our approach consists in assuming that, if structural effects are neglected, $H_m^E$ is then the result of three contributions [64,83]. Two positive ones, $\Delta H_{OH-OH}$, $\Delta H_{X-X}$, which arise, respectively, from the disruption of alkanol-alkanol and X-X interactions along the mixing process, and a negative third contribution, $\Delta H_{OH-X}$, due to the new OH---X interactions created upon mixing. Therefore [38,84-86]:

$$H_m^E = \Delta H_{OH-OH} + \Delta H_{X-X} + \Delta H_{OH-X} \tag{12}$$

It is possible to conduct an evaluation of $\Delta H_{OH-X}$ extending the equation (12) to $x_1 \to 0$ [38, 86,87]. Then, $\Delta H_{OH-OH}$ and $\Delta H_{X-X}$ can be replaced by $H_{m1}^{E,\infty}$ (partial excess molar enthalpy at infinite dilution of the first component) of 1-alkanol or polar compound + heptane systems. Thus,

$$\Delta H_{OH-X} = H_{m1}^{E,\infty}(1-\text{alkanol} + \text{polar compound})$$

$$-H_{m1}^{E,\infty}(1-\text{alkanol} + \text{heptane}) - H_{m1}^{E,\infty}(\text{polar compound} + \text{heptane}) \tag{13}$$

As in other applications, we have assumed that, for 1-alkanol + *n*-alkane systems, the $H_{m1}^{E,\infty}$ value is independent of the alcohol, which is a very common approach [21,70,87-90]. The value used in this work is the same as in previous studies [38,82], $H_{m1}^{E,\infty} = 23.2$ kJ·mol$^{-1}$ [91-93]. The determination of the $H_{m1}^{E,\infty}$ for DMSO or sulfolane + alkane mixtures is rather difficult since calorimetric data are not available due to the extremely large miscibility gaps of these solutions (see above). In the case of DMSO + alkane, the enthalpy of the DMSO-DMSO interactions have been evaluated, in the framework of the ERAS model, to be equal $-25$ kJ mol$^{-1}$, i.e., $H_{m1}^{E,\infty} = 25$ kJ mol$^{-1}$ [94]. The corresponding value for sulfolane (32 kJ mol$^{-1}$) has been obtained from activity coefficients at infinite dilution of this compound for hexane mixtures at the temperature range (334.6-341.4) K [95]. In spite of the shortcomings involved along calculations, results are still meaningful since they have been obtained in similar way that those previously reported. Figure 9 shows the variation of $\Delta H_{OH-X}$ with $n_{OH}$ for the systems 1-alkanol + NM, or + EtN, or + DMSO, or + SULF, or + NTBz, or + BzCN. For a fixed 1-alkanol, interactions between unlike molecules become weaker in the order: DMSO $\approx$ SULF > EtN > NM > BzCN > NTBz. For a given polar compound, these interactions are weakened when $n_{OH}$ increases, probably because the OH group becomes more sterically hindered.

*5.2.3 Excess enthalpies and excess molar entropies*

We describe now some common features of the systems under consideration. (i) $H_m^E$ values are usually large and positive (Table 2). For example, $H_m^E$(1-propanol)/J·mol$^{-1}$ = 1911 (NM) [96]; 1829 (EtN) [97]. That is, the main contribution to $H_m^E$ arises from the breaking of the interactions between like molecules. The methanol + DMSO system is an exception since $H_m^E = -391$ J·mol$^{-1}$ [98], and the contribution to this excess function from the interactions between unlike molecules is here dominant. (ii) The $H_m^E$ curves of mixtures involving longer 1-alkanols are skewed to higher mole fractions of the alcohol (see, eg. [81,96]). These curves are nearly symmetrical for systems with $n_{OH} = 1-3$ [81,96]. (iii) Both $H_m^E$ and $n_{OH}$ increase in line (Table 2). (iv) Except for the methanol + NTBz system, $TS_m^E$ values are positive (Table 2). These are typical features of mixtures where self-association or solvation effects are not relevant. Systems where the alcohol self-association is determinant show much lower $H_m^E$ values, which decrease when the alcohol size is increased (e.g., 1-alkanol + alkane [41]), or which remain nearly constant for the solutions containing longer 1-alkanols (e.g., 1-alkanol + dibutylether [99]). The $H_m^E$ curves are shifted towards low mole fractions of the 1-alkanol and the $TS_m^E$ curves show negative values over almost the entire composition range, and positive values at low alcohol concentrations [92]. In the case of the ethanol + hexane mixture, $H_m^E = 548$ [100];

$G_m^E$ = 1374 [92] and $TS_m^E$ = −826 (all results in J·mol$^{-1}$). Values of $TS_m^E$ and $n_{OH}$ also increase in line. These features are explained in terms of the alcohol self-association [93]. The decrease of alcohol self-association when $n_{OH}$ is increased leads to increasing $TS_m^E$ values as the alcohol network is more easily disrupted. Values of $TS_m^E$ are positive at low alcohol concentrations as then interactions between alcohol molecules are more easily broken, an effect that also determines the concentration dependence for $H_m^E$.

The $H_m^E$ increase with $n_{OH}$ (Table 2), along a homologous series, can be explained taking into account that, at this condition, $\Delta H_{OH-X}$ increases (is less negative) and $\delta_{21}$ decreases, i.e., the interactions between unlike molecules become weaker and the number of such interactions is lower. It is remarkable that, within good approximation, a linear dependence exist between both $H_m^E$ and $\Delta H_{OH-X}$ for mixtures including EtN, NM, BzCN or NTBz or DMSO and $n_{OH}$ = 1-4. Mixtures with longer 1-alkanols and EtN or DMSO are less sensitive to $\Delta H_{OH-X}$. (Table 7). For the homologous series including NM or BzCN, $H_m^E$ also changes linearly with the $\delta_{21}$ minimum value (Table 7). Results are somewhat poorer for solutions with NM or DMSO. For a given 1-alkanol, $H_m^E$ values decrease when NM is replaced by EtN or NTBz is replaced by BzCN. This is due to, for nitriles, the contribution to $H_m^E$ from the disruption of solvent-solvent interactions is lower (weaker interactions), and the corresponding contribution to $H_m^E$ from the creation of interactions between unlike molecules is higher in absolute vale (stronger interactions). Regarding the latter point, it should be also kept in mind that nitriles ar more preferred around a central 1-alcohol molecule than nitroalkanes. The very large positive $H_m^E$ values of SULF systems may be related to the breaking of the very strong interactions between sulfolane molecules.

On the other hand, $TS_m^E$ values also increase with $n_{OH}$ along a homologous series (Table 2). This remarks that association/solvation effects become progressively of minor importance, particularly at higher $n_{OH}$ values. In fact, at 298.15 K, the system temperature of solutions with, e.g, NM or EtN and longer 1-alkanols is closer to the corresponding UCST. We compare now $TS_m^E$ values for different homologous series. In systems with a fixed 1-alkanol, $TS_m^E$ values decrease when EtN is replaced by NM. This is due to the $G_m^E$ increase, that is the $G_m^M$ increase, when EtN is replaced by NM is higher than the corresponding $H_m^E$ increase (Table 2). This suggests that the decrease in the number of interactions between unlike molecules when replacing EtN by NM is accompanied by a lower $H_m^E$ increase than that which one could expect, probably because NM-NM interactions are stronger and it is more difficult to break them. DMSO systems including longer 1-alkanols are characterized by low and positive $G_m^E$ values, whereas $H_m^E$ and $TS_m^E$ values are large and positive (enthalpic-entropic compensation, Table 2). This merely shows that such 1-alkanols are good breakers of the strong DMSO-DMSO interactions. Thus, although a large number of interactions between unlike molecules exist, the contribution to $H_m^E$ from the breaking of interactions between like molecules is largely dominant. For systems with $n_{OH}$ =1-4 and DMSO, interactions between unlike molecules are more probable ($G_m^E < 0$, and consequently more negative $G_m^M$ values). However, except for the methanol mixture, the contribution to $H_m^E$ related to the disruption interactions between like molecules is still dominant.

*5.2.4   Excess molar heat capacities at constant pressure*

Direct calorimetric $C_{pm}^E$ measurements are scarce and the same occurs for $H_m^E$ values at different temperatures for the studied systems. In spite of this, some interesting statements can be given with regards to the temperature dependence of $H_m^E$. (i) For NM systems, $C_{p,m}^E$/ J·mol$^{-1}$·K$^{-1}$ = 9.3 ($n_{OH}$ =1) [101]; 16.6 ($n_{OH}$ = 3) [102]; 20.6 ($n_{OH}$ = 4) [36]. The large values measured for $n_{OH}$ = 3,4 are due to the proximity of the UCST. Accordingly, $C_{p,m}^E$ decreases when the temperature is increased [36,102]. The result for the methanol system suggests that self-association effects may have importance. Note that for the ethanol + heptane mixture, $C_{p,m}^E$ = 11.7 J·mol$^{-1}$·K$^{-1}$ [103]. For EtN solutions, we have estimated $C_{p,m}^E$ values from $H_m^E$ at different temperatures. Thus, $C_{p,m}^E$/J·mol$^{-1}$·K$^{-1}$ = 5.3 ($n_{OH}$ = 1); 7.4 ($n_{OH}$ = 2) [104]; 7.7 ($n_{OH}$ = 5); 13.7 ($n_{OH}$ = 6) [105]. This is a similar behavior to that observed for NM mixtures, as $C_{p,m}^E$ increases sharply when the system temperature is closer to the UCST, that is, for the longer 1-alkanols. For DMSO mixtures, the available data are in contradiction. For the 1-propanol mixture, the directly measured $C_{p,m}^E$ value is $-10.1$ J·mol$^{-1}$·K$^{-1}$, whereas experimental $H_m^E$ results change much more slowly with temperature: 1.6 J·mol$^{-1}$·K$^{-1}$ [106].

*5.2.5   The $G_m^E$ vs. $H_m^E$ diagram*

We briefly summarize some important features of this type of diagrams. More details can be found elsewhere [107-109]. (i) $G_m^E = H_m^E/2$ is the dividing line between positive and negative values of $C_{pm}^{E}$. Systems below this line show negative $C_{pm}^{E}$ values. The line $G_m^E = H_m^E$ divides the diagram in two parts with a different sign for $TS_m^E$. (ii) In the first quarter of the plot, non-associated mixtures are situated between the lines $G_m^E = H_m^E/3$ and $G_m^E = H_m^E/2$. Such solutions are characterized by $C_{pm}^{E} < 0$ and $TS_m^E > 0$. For example, for the cyclohexane + hexane system, $G_m^E$ = 101 [110]; $H_m^E$ = 230 [111]; $TS_m^E$ = 129, all results in J·mol$^{-1}$ and $C_{pm}^{E}$ = −1.39 J·mol$^{-1}$·K$^{-1}$ [112]. (iii) Mixtures with self-associated compounds are encountered in the region well above from the line $G_m^E = H_m^E$. These solutions have $C_{pm}^{E} > 0$ and $TS_m^E < 0$ (see above). (iv) In the region between the lines $G_m^E = H_m^E/2$ and $G_m^E = H_m^E$, we find mixtures such as *n*-alkanone, or linear organic carbonate, or *N*-methylpyrrolidone + alkane, i.e. systems characterized by dipolar interactions. Thus, for the dimethyl carbonate + heptane mixture, $G_m^E$ = 1156 J·mol$^{-1}$, value obtained using DISQUAC interaction parameters from the literature [113]; $H_m^E$ = 1988 J·mol$^{-1}$ [114]; $TS_m^E$ = 832 J·mol$^{-1}$ and $C_{pm}^{E}$ = 2.83 J·mol$^{-1}$·K$^{-1}$ [115]. (v) If solvation exists, then the solutions are situated in the third quarter of the diagram. This is the case of the 2-propanone + CHCl$_3$ mixture, with $G_m^E$ = −605 [116] and $H_m^E$ = −1972 [117] J·mol$^{-1}$. We note that many of the systems under consideration are placed in the first quarter of the diagram (Figure 10), between the lines $G_m^E = H_m^E/2$ and $G_m^E = H_m^E$. Therefore, they have $C_{pm}^{E} > 0$ and $TS_m^E > 0$. One should say that dipolar interactions are here rather relevant. Interestingly, some 1-alkanol + EtN mixtures are located close to the $G_m^E = H_m^E/2$ line, i.e, close to the region of non-associated systems. This is consistent with our previous investigation of these mixtures in the framework of the Flory model, that shows that orientational effects are of minor importance for systems with ethanol 1-propanol or 1-butanol (see above). The methanol + NTBz system is situated above but very close to the line $G_m^E = H_m^E$, indicating that self-association effects are scarcely relevant. The methanol + DMSO mixture is encountered in the region where solvation effects are determinant. The remainder DMSO systems are in the first and fourth quarters and have the larger positive $TS_m^E$ values. In large extent, mixing is determined for such solutions by entropic effects.

### 5.2.6  *Structural effects and magnitudes at constant volume*

Interestingly, many of the present systems show different signs for $H_m^E$ and $V_m^E$ functions (Table 2), which indicates the existence of structural effects [118,119]. An extreme case is that for 1-alkanol + SULF mixtures, with very large positive $H_m^E$/J·mol$^{-1}$ values at 303.15 K (1551

($n_{OH}$ = 1); 1971 ($n_{OH}$ = 2)) [120] and very low negative $V_m^E$/cm$^3$·mol$^{-1}$ values at the same temperature ($-0.826$ ($n_{OH}$ = 1); $-0.697$ ($n_{OH}$ = 2)) [121]. Other mixtures, e.g., the 1-hexanol + EtN system, are characterized by large positive $H_m^E$ values (2313 J·mol$^{-1}$ [105]) and low positive $V_m^E$ values (0.205 cm$^3$·mol$^{-1}$ [122]), which is also indicative of the existence of structural effects. As a rule, both $H_m^E$ and $V_m^E$ values increase in line along a given homologous series (Table 2), which suggests the $V_m^E$ variation is closely related to that of the interactional effects.

It is well known that $H_m^E$ values are not entirely determined by interactional effects, but also to structural effects [64,83]. The former are more properly considered using $U_{Vm}^E$, the excess internal energy at constant volume. If terms of higher order in $V_m^E$ are neglected, $U_{Vm}^E$ can be written as [64,83]:

$$U_{Vm}^E = H_m^E - \frac{\alpha_p}{\kappa_T} T V_m^E \qquad (14)$$

where $\frac{\alpha_p}{\kappa_T} T V_m^E$ is the equation of state (eos) contribution to $H_m^E$, and $\alpha_p$ is the isobaric thermal expansion coefficient of the mixture. In this work, $\alpha_p$ and $\kappa_T$ were determined assuming ideal behavior ($F = \varphi_1 F_1 + \varphi_2 F_2$; $F_i$ is the property of the pure compound $i$) for these magnitudes. Structural effects have a weak impact on $H_m^E$ data for many of the investigated systems, since $H_m^E$ and $U_{Vm}^E$ results do not differ substantially (Table 2). Notable exceptions are encountered for the mixtures (between parenthesis are indicated absolute differences in % between $H_m^E$ and $U_{Vm}^E$ values): methanol + SULF (20.6 %), or + NTBz (14.8%), or + BzCN (13.2%), or + DMSO (58.6%). Interestingly, molecular dynamic simulations predict positive $U_{Vm}^E$ values for the methanol + DMSO system (300 J·mol$^{-1}$) [123], which is very different to the experimental value given in Table 2 ($-162$ J·mol$^{-1}$). The $G_m^E$ vs. $U_{Vm}^E$ diagram is very similar to the $G_m^E$ vs. $H_m^E$ diagram, and the discussion conducted above is still valid. The main change affects to entropies as $H_m^E - U_{Vm}^E \approx TS_m^E - TS_{Vm}^E$ [64]. Thus, a change of this magnitude is produced for the mehanol + NTBz system since $TS_m^E = -147$ and $TS_{Vm}^E = 17$ (values in J·mol$^{-1}$)

*5.2.7 Dielectric constants and Kirkwood's correlation factor*

The values of excess permittivities, $\varepsilon_r^E$, at $\varphi_1 = 0.5$ are collected in Table 3, and were determined, from the original data, according to the equation [124]:

$$\varepsilon_r^E = \varepsilon_r - \varphi_1 \varepsilon_{r1} - \varphi_2 \varepsilon_{r2} \tag{15}$$

($\varepsilon_{ri}$ is the permittivity of pure compound i). The methanol + EtN, or + SULF, or + BzCN, or + DMSO systems are characterized by $\varepsilon_r^E > 0$. In such cases, the interactions between unlike molecules lead to the formation of multimers of higher effective dipole moments than those of the pure compounds and this positive contribution to $\varepsilon_r^E$ is prevalent over those arising from the disruption of interactions between like molecules [125,126]. For the remainder systems under consideration, $\varepsilon_r^E < 0$, and the contribution to $\varepsilon_r^E$ from the breaking of the alcohol network and of the dipolar interactions between solvent molecules is dominant [125,127-130]. It is to be noted that the dependence of $\varepsilon_r^E$ for 1-alkanol + EtN, or + NTBz with $n_{OH}$ is similar to that observed for other mixtures such as 1-alkanol + cyclohexane, or + cyclohexylamine, or + dipropylether [125]. This behaviour has been explained in terms of the weaker and lower self-association of longer 1-alkanols [125] and seems to be a general trend. Although $\varepsilon_r^E$ data for solutions with EtN or NM and a given 1-alkanol are at different temperatures, it is possible to conclude that cooperative effects which lead to a more effective polarization of the mixture are more relevant in EtN solutions since $\varepsilon_r^E$(EtN) is much higher than $\varepsilon_r^E$(NM). The same occurs for BzCN systems with respect to those including NTBz. That is, alkanol-nitrile interactions contribute more positively to $\varepsilon_r^E$ than alkanol-nitroalkane interactions. The large and negative $\varepsilon_r^E$ values of 1-propanol, or 1-butanol + NTBz mixtures, compared with the results for NM mixtures, are remarkable. They reveal that the structure of the mixture compounds is largely broken upon mixing. This may due to NM-NM interactions are stronger than those between NTBz molecules and to the aromatic compounds are better breakers of the alkanol self-association. On the other hand, cooperative effects are much important in SULF systems than in NM solutions, which is consistent with the results obtained from the Kirkwood-Buff integrals. A surprising result is encountered for DMSO mixtures: $\varepsilon_r^E$(methanol) = 4.04 [80] < $\varepsilon_r^E$(1-propanol) = 6.19 [131]. This unusual behaviour might be supported by $^1$NMR spectral studies that show that the most stable complex is of the form DMSO·3propanol [131] in the 1-propanol solution, while is of the type DMSO·2methanol in the methanol mixture. Nevertheless, it must be underlined that large discrepancies on $\varepsilon_r^E$ values for the 1-propanol + DMSO system exist in the literature. Thus, $\varepsilon_r^E = -0.44$ [132,133] or $-4.9$ [134].

Inspection of $g_K$ results listed in Table 3 allows state that the considered mixtures are rather unstructured in terms of the dielectric polarization. In fact, at $\varphi_1 = 0.5$, $g_K$ values are slightly larger than 1, or very close to 1. In addition, along a homologous series, when $n_{OH}$ increases, $g_K(\varphi_1)$ curves have progressively a wider region where $g_K$ values remain close to 1 (Figure 6). That is, solutions become more unstructured, as interactions between unlike molecules are then less relevant. For the sake of comparison, Figure 6 also includes the $g_K(\varphi_1)$ curve for the methanol + hexylamine mixture. Its concentration dependence is very different, showing a sharp $g_K$ increase at low $\varphi_1$ values, which indicates that, in that region, interactions between unlike molecules largely contribute to the mixture polarization.

*5.5    Dynamic viscosities*

We discuss now the values of viscosity deviations from the linear behaviour, calculated using the equation:

$$\Delta \eta = \eta - x_1 \eta_1 + x_2 \eta_2 \tag{16}$$

where $\eta_i$ stands for the dynamic viscosity of component i. Results given below are at equimolar composition. Firstly, it must be remarked that the $\Delta \eta$ values listed in Table 3 are negative. They can be explained assuming that the mixing process leads to a higher fluidization of the solution due to the breaking of alcohol self-association and of dipolar interactions between solvent molecules. These negative contributions to $\Delta \eta$ are prevalent over the positive contribution related to the interactions between unlike molecules. On the other hand, for mixtures with a given solvent, $\Delta \eta$ values decrease with the increasing of $n_{OH}$, since alkanol-solvent interactions become then less probable. Accordingly with this statement, larger $\Delta \eta$ values are encountered for methanol systems. An exception is the sulfolane solution, which is consistent with the very large and positive $H_m^E$ value of this system. Viscosity and density data can be used to determine the enthalpy, $\Delta H^*$, and entropy, $\Delta S^*$, of activation of viscous flow on the basis of the Eyring's theory [135-137]. The starting equation is [138,139]:

$$\ln \frac{\eta V_m}{h N_A} = \frac{\Delta H^*}{RT} - \frac{\Delta S^*}{R} \qquad (17)$$

The plots of ln($\eta V/hNA$) vs. $1/T$ give a straight line for each mixture and $\Delta H^*$, and $\Delta S^*$ can be estimated from its slope and intercept. We have calculated the values for the methanol + DMSO, or + NM, or + EtN systems at equimolar composition and 298.15 K using data from the literature [140-142]. The results are: $\Delta H^*$/kJ·mol$^{-1}$ = 13.9 (DMSO), 9.4 (NM), 8.1 (EtN) and $\Delta S^*$/kJ·mol$^{-1}$·K$^{-1}$ = 5 (DMSO), $-2$ (NM), $-3.9$(EtN). It is clear that the activation process from the initial state to the transition state at a given composition is mainly determined by enthalpic effects. If one takes into account the $\eta_i$/mPa·s values of pure solvents in methanol systems (1.991 (DMSO), 0.614 (NM), 0.341 (EtN)) [1], it seems that $\Delta H^*$ is largely dependent on such values. Other effects are, of course, present. It is pertinent to compare results for methanol + DMSO, or + cyclohexylamine mixtures since both solvents have very similar $\eta_i$ values (1.908 mPa·s for cylohexylamine [143]). For the amine solution, $\Delta H^* = 17.6$ kJ·mol$^{-1}$; $\Delta S^* = 10.2$ J·mol$^{-1}$·K$^{-1}$ [143]. These results suggest the importance of interactional effects on $\Delta H^*$ values. We must remark the large difference existing between the $H_m^E$/J·mol$^{-1}$ values of these systems: $-3248$ (cyclohexylamine) [144], $-391$ (DMSO) [98]. It seems that $\Delta H^*$ increases when alkanol-solvent interactions become stronger. Nevertheless, an extension of the available database, including new accurate viscosity measurements, is required to investigate this matter in detail, as size and shape effects are also relevant when analyzing viscosity data [145].

### 6. Conclusions

Many of the considered systems are located between the lines $G_m^E = H_m^E/2$ and $G_m^E = H_m^E$, and consequently are characterized by dipolar interactions, showing positive values for $H_m^E$, $C_{pm}^E$ and $TS_m^E$. For systems with a given 1-alkanol, interactions between unlike molecules become weaker in the order: DMSO $\approx$ SULF > EtN > NM > BzCN > NTBz; and in systems with a fixed solvent they are weakened when the 1-alkanol size increases. The studied mixtures also show strong structural effects. In fact, $H_m^E$, and $U_{Vm}^E$ results do not largely differ, except for a few systems (58.6%. for the methanol + DMSO system). The impact of structural effects is larger for excess entropies. Calculations on Kirkwood-Buff integrals reveal that systems containing NM, or NTBz, or BzCN or SULF or the 1-octanol + EtN mixture are characterized by interactions between like molecules. In contrast, local mole fractions are very similar to the bulk ones for DMSO systems or for mixtures including EtN and shorter 1-alkanols. In such cases, the mixture structure can be ascribed to the existence of orientational effects. On the other hand, our results indicate that nitriles are more preferred than nitroalkanes around a central alcohol molecule. It has been also shown, from the $g_K$ values obtained in this work, that the systems are rather unstructured, and that this trend becomes more relevant when the chain length of the 1-alkanol increases in solutions with a given solvent.

## 7. List of symbols

| | |
|---|---|
| $C_p$ | heat capacity at constant pressure |
| $\Delta H$ | enthalpy of interaction (equation 13) |
| $\Delta H_{vap}$ | standard enthalpy of vaporization |
| $g_K$ | Kirkwood's correlation factor (equation 6) |
| $G$ | Gibbs energy |
| $G_{ij}$ | Kirkwood-Buff integral (equations 1-3) |
| $H$ | enthalpy |
| $n_{OH}$ | number of C atoms in 1-alkanol |
| $S$ | entropy |
| $S_V$ | entropy at constant volume |
| $T$ | temperature |
| $U_V$ | internal energy at constant volume |
| $V$ | volume |
| $x$ | mole fraction in liquid phase |

*Greek letters*

| | |
|---|---|
| $\alpha_P$ | isobaric thermal expansion coefficient |
| $\varepsilon_r$ | relative permittivity |
| $\varphi$ | volume fraction |
| $\eta$ | dynamic viscosity |
| $\kappa_T$ | isothermal compressibility |
| $\mu$ | dipole moment |
| $\bar{\mu}$ | effective dipole moment (equation 10) |
| $\sigma_r$ | relative standard deviation (equation 11) |
| $X_{12}$ | interaction parameter in the Flory model |

**Superscripts**

| | |
|---|---|
| E | excess property |

**Subscripts**

| | |
|---|---|
| i,j | compound in the mixture, (i, j =1,2) |
| m | molar property |

## 8. References

[1] J.A. Riddick, W.B. Bunger, T.K. Sakano, Organic Solvents: Physical Properties and Methods of Purification, fourth ed., Wiley, New York, 1986.

[2] A.A. Gaile, G.D. Zalishchevskii, A.S. Erzhenkov, E.A. Kayfazhyan, L.L. Koldobskaya, Russ. J. Appl. Chem. 80 (2007) 591-594.

[3] P.J. Bailes, Chem. Ind. 15 (1977) 69-73.

[4] F.F. Fleming, L. Yao, P.C. Ravikumar, L. Funk, B.C. Shook,. J. Med. Chem. 25 (2010) 7902-7917.

[5] E. Boyer, K.K. Kuo, Proceed. Comb. Inst. 31 (2007) 2045-2053

[6] T. Radnai, S. Ishiguro, H.Ohtaki, Chem. Phys. Lett 159 (1989) 532-537.

[7] R. S. Catalotti, G. Paliani, L. Mariani, S. Santini, and M. G. Giorgini, J. Phys. Chem. 96 (1992) 2961-2964.

[8] T. Megyes, S. Balint, T. Grosz, T. Radnai, I. Bakó, L. Almásy, J. Chem. Phys. 126 (2007) 164507-164511.

[9] M. Della Monica, L. Jannelli, U. Lamanna, J. Phys. Chem. 72 (1968) 1068-1071.

[10] M. Pansini, L. Jannelli, J. Chem. Eng. Data, 31 (1986) 157-160.

[11] L. Jannelli, A. López, S. Saiello, J. Chem. Eng. Data 25 (1980) 259-263.

[12] L. Jannelli, A. Azzi, A. López, S. Saiello, J. Chem. Eng. Data 25 (1980) 77-79.

[13] J.A. González, U. Domanska, Phys. Chem. Chem. Phys. 3 (2001) 1034-1042.

[14] J.A. González, S. Villa, N. Riesco, I. García de la Fuente, J.C. Cobos, Phys. Chem. Liq. 41 (2003) 583-597.

[15] J.A. González, F. Hevia, A. Cobos, I. García de la Fuente, C. Alonso-Tristán. Thermochim. Acta 605 (2015) 121-129.

[16] J.A. González, I. García de la Fuente, J.C. Cobos, C. Alonso-Tristán, L.F. Sanz, Ind. Eng. Chem. Res. 54 (2015) 550-559.

[17] J.A. González, F. Hevia, L.F. Sanz, I. García de la Fuente, C. Alonso-Tristán, Fluid Phase Equilib. 471 (2018) 24-39.

[18]  J.A. González, C. Alonso-Tristán, F. Hevia, I. García de la Fuente, L.F. Sanz, J. Chem. Thermodyn. 116 (2018) 259-272.

[19]  H.V. Kehiaian, Fluid Phase Equilib. 13 (1983) 243-252.

[20]  J.A. González, I. García de la Fuente, J.C. Cobos. Correlation and prediction of excess molar enthalpies using DISQUAC in: E. Wilhelm, T.M. Letcher (Eds.), Enthalpy and Internal Energy: Liquids, Solutions and Vapours, Royal Society of Chemistry, Croydon 2017.

[21]  A. Heintz, Ber. Bunsenges. Phys. Chem. 89 (1985) 172-181.

[22]  P.J. Flory, J. Am. Chem. Soc. 87 (1965) 1833-1838.

[23]  J.C. Cobos, Fluid Phase Equilib. 133 (1997) 105-127.

[24]  W.H. Young, Rep. Prog. Phys. 55 (1992) 1769-1853.

[25]  J.G. Kirkwood, F.P. Buff, J. Chem. Phys. 19 (1951) 774-777.

[26]  A. Ben-Naim, J. Chem. Phys. 67 (1977) 4884-4889.

[27]  J.C.R. Reis, T.P. Iglesias, Phys. Chem. Chem. Phys. 13 (2011) 10670-10680.

[28]  H. Fröhlich, Theory of Dielectrics, Clarendon Press, Oxford, 1958.

[29]  A. Chelkowski, Dielectric Physics, Elsevier, Amsterdam, 1980.

[30]  E. Matteoli, J. Phys. Chem. B. 101 (1997) 9800-9810.

[31]  J. Zielkiewicz, Phys. Chem. Chem. Phys. 5 (2003) 1619-1630.

[32]  Y. Marcus, J. Solution Chem. 21 (1992) 1217-1230.

[33]  M. El-Hefnawy, K. Shameshima, T. Matsushita, R. Tanaka, J. Solution Chem. 34 (2005) 43-69.

[34]  J. Zielkiewicz, J. Chem. Soc., Faraday Trans. 94 (1998) 1713-1719.

[35]  E. Matteoli, L. Lepori, J. Chem. Phys. 80 (1984) 2856-2863.

[36]  C.A. Cerdeiriña, C.A. Tovar, D. González, E. Carballo, L. Romaní, Fluid Phase Equilib. 179 (2001) 101-115.

[37]  M.S. Bakshi, J. Singh, H. Kaur, S.T. Ahmad, G. Kaur, J. Chem. Eng. Data 41 (1996) 1459-1461.

[38]  J.A. González, I. Mozo, I. García de la Fuente, J.C. Cobos, N. Riesco, J. Chem. Thermodyn. 40 (2008) 1495-1508.

[39]  S.B. Aznarez, M.A. Postigo, J. Solution Chem. 27 (1998) 1045-1053.

[40]  M.A. Motin, M. Azhar Ali, Sh. Sultana, Phys. Chem. Liq. 45 (2007) 221-229.

[41]  J.A. González, I. García de la Fuente, J.C. Cobos, C. Casanova, Ber. Bunsenges. Phys. Chem. 95 (1991) 1658-1668.

[42]  U. Domanska, J.A. González, Fluid Phase Equilib. 119 (1996) 131-151.

[43]  B. Marongiu, B. Pittau, S. Porcedda, Thermochim. Acta, 221 (1993) 143-162.

[44] B. Marongiu, S. Porcedda, H.V. Kehiaian, Fluid Phase Equilib. 87 (1993) 115-131.

[45] J.A. González, I. García de la Fuente, J.C. Cobos, C. Casanova, J. Solution Chem. 23 (1994) 399-420.

[46] U. Domanska, J.A. González, Fluid Phase Equilib. 123 (1996) 167-187.

[47] J.A. González, I. García de la Fuente, J.C. Cobos, C. Casanova, Fluid Phase Equilib. 93 (1994) 1-22.

[48] A. Cannas, B. Marongiu, S. Porcedda, Thermochim. Acta 311 (1998) 1-19.

[49] E.R. Thomas, B.A. Newman, T.C. Long, D.A. Wood, C.A. Eckert, J. Chem. Eng. Data 27 (1982) 399-405.

[50] J.C.R. Reis, I.M.S. Lampreia, Â.F.S. Santos, M.L.C.J. Moita, G. Douhéret, ChemPhysChem, 11 (2010) 3722-3733.

[51] A Rodríguez, J. Canosa, J. Tojo, J. Chem. Eng. Data 46 (2001) 1506-1515.

[52] R.K. Shukla, P. Misra, S. Sharma, N. Tomar, P. Jain, J. Iran. Chem. Soc. 9 (2012) 1033-1043.

[53] A. Piñeiro, P. Brocos, A. Amigo, M. Pintos, R. Bravo, J. Solution Chem. 31 (2002) 369-380.

[54] P.S. Nikam, B.S. Jagdale, A.B. Sawant, M. Hasan, J. Chem. Eng. Data 45 (2000) 214-218.

[55] N.G. Tsierkezos, M.M. Palaiologou, I.E. Molinou, J. Chem. Eng. Data 45 (2000) 272-275.

[56] M. Ko, J. Im, J.-Y. Sung, H. Kim, J. Chem. Eng. Data 52 (2007) 1464-1467.

[57] A. Nissema, S. Sarkela, Suomi. Kemistil B46 (1973) 58 (cf. [58]).

[58] J.M. Sorensen, W. Arlt, Liquid-liquid Equilibrium Data Collection. Binary Systems. Chemistry Data Series, Vol. V. Part 1. DECHEMA, Frankfurt/Main, Germany, 1979.

[59] G. Spinolo. Int. DATA Ser., Sel. Data Mixtures, Ser. A 1 (1981) 21-29.

[60] J.B. Ott, J.B. Purdy, B.J. Neely, R.A. Harris, J. Chem. Thermodyn. 20 (1988) 1079-1087.

[61] H.V. Kehiaian, M.R. Tiné, L. Lepori, E. Matteoli, B. Marongiu, Fluid Phase Equilib. 46 (1989) 131-177.

[62] J.A. González, I. García de la Fuente, J.C. Cobos, Fluid Phase Equilib. 154 (1999) 11-31.

[63] J.A. González, I. Alonso, C. Alonso-Tristán, I. García de la Fuente, J.C. Cobos, J. Chem. Thermodyn. 56 (2013) 89-98.

[64] J.S. Rowlinson, F.L. Swinton, Liquids and Liquid Mixtures, third ed., Butterworths, London, 1982.

[65] J.A. González, I. García de la Fuente, J.C. Cobos, Fluid Phase Equilib. 168 (2000) 31-58.

[66]   E. Wilhelm, A. Laínez, J.-P.E. Grolier, Fluid Phase Equilib. 49 (1989) 233-250.

[67]   J.A. González, I. Mozo, I. García de la Fuente, J.C. Cobos, V. Durov. Fluid Phase Equilib. 245 (2006) 168-184.

[68]   T. Kimura, K. Suzuki, S. Takagi, Fluid Phase Equilib. 136 (1997) 269-278.

[69]   V.P. Sazonov, N.I. Lisov, N.V. Sazonov, J. Chem. Eng. Data 47 (2002) 599-602.

[70]   J. Zielkiewicz, J. Phys. Chem. 99 (1995) 3357-3364.

[71]   A. Ben Naim, Cell Biophys. 12 (1988) 255-269.

[72]   K.E. Newman, J. Chem. Soc., Faraday Trans. 1, 84 (1988) 1387-1391.

[73]   V.P. Sazonov, V.V. Filippov, Izv. Vysh. Ucheb. Zaved Khim. Khim. Tekhnol. 18 (1975) 222 (cf. [58]).

[74]   J. Zielkiewicz, Phys. Chem. Chem. Phys. 2 (2000) 2925-2932.

[75]   J. Zielkiewicz, Phys. Chem. Chem. Phys. 2 (2003) 3193-3201.

[76]   J.A. González, J.C. Cobos, I. García de la Fuente, Fluid Phase Equilib. 224 (2004) 169-183.

[77]   J.A. González, I. García de la Fuente, I. Mozo, J.C. Cobos, N. Riesco, Ind. Eng. Chem. Res. 47 (2008) 1729-1737

[78]   J.A. González, N. Riesco, I. Mozo, I. García de la Fuente, J.C. Cobos, Ind. Eng. Chem. Res. 48 (2009) 7417-7429.

[79]   Y. Marcus, J. Solution Chem. 35 (2006) 251-277.

[80]   S.J. Romanowski, C.M. Kinart, W.J. Kinart, J. Chem. Faraday Trans. 91 (1995) 65-70.

[81]   T. Kimura, T. Matsushita, M. Momoki, H. Mizuno, N. Kanbayashi, T. Kamiyama, M. Fujisawa, S. Takagi, Y. Toshiyasu, Thermochim. Acta 424 (2004) 83-90.

[82]   J.A. González, A. Mediavilla, I. García de la Fuente, J.C. Cobos, J. Chem. Thermodyn. 59 (2013) 195-208.

[83]   H. Kalali, F. Kohler, P. Svejda, Fluid Phase Equilib. 20 (1985) 75-80.

[84]   T.M. Letcher, U.P. Govender, J. Chem. Eng. Data 40 (1995) 1097-1100.

[85]   E. Calvo, P. Brocos, A. Piñeiro, M. Pintos, A. Amigo, R. Bravo, A.H. Roux, G. Roux-Desgranges, J. Chem. Eng. Data 44 (1999) 948-954.

[86]   T.M. Letcher, B.C. Bricknell, J. Chem. Eng. Data 41 (1996) 166-169.

[87]   J.A. González, I. García de la Fuente, J.C. Cobos, Fluid Phase Equilib. 301 (2011) 145-155.

[88]   V. Brandani, J.M. Prausnitz, Fluid Phase Equilib. 7 (1981) 233-257.

[89]   A. Liu, F. Kohler, L. Karrer, J. Gaube, P.A. Spelluci, Pure & Appl. Chem. 61 (1989) 1441-1452.

[90]   H. Renon, J.M. Prausnitz, Chem. Eng. Sci. 22 (1967) 299-307.

[91]   R.H. Stokes, C. Burfitt, J. Chem. Thermodyn. 5 (1973) 623-631.

[92]  S.J. O'Shea, R.H. Stokes, J. Chem. Thermodyn. 18 (1986) 691-696

[93]  H.C. Van Ness, J. Van Winkle, H.H. Richtol, H.B Hollinger, J. Phys. Chem. 71 (1967) 1483-1494.

[94]  T.M. Letcher, P.G. Whitehead, J. Chem. Thermodyn. 31 (1999) 1537-1549.

[95]  M. Mukhopadhyay, A.S. Pathak, J. Chem. Eng. Data 31 (1986) 148-152.

[96]  I. Hammerl, A. Feinbube, K. Herkener, H.-J. Bittrich, Z. Phys. Chem. (Leipzig) 271 (1990) 1133-1138.

[97]  I. Nagata, K. Tamura, J. Chem. Thermodyn. 20 (1988) 87-93.

[98]  T. Kimura, T. Morikuni, T. Chanoki, S. Takagi, Netsu Soktei 17 (1990) 67-72.

[99]  I. Mozo, I. García de la Fuente, J.A. González, J.C. Cobos, J. Chem. Thermodyn. 42 (2010) 17-22.

[100] L. Wang, G.C. Benson, B. C.-Y. Lu, J. Chem. Thermodyn. 24 (1992) 1135-1143.

[101] H. Piekarski, A. Pietrzak, D. Waliszewski, J. Mol. Liq. 121 (2005) 41-45.

[102] C.A. Cerdeiriña, C.A. Tovar, J. Troncoso, E. Carballo, L. Romaní, Fluid Phase Equilib. 157 (1999) 93-102

[103] R. Tanaka, S. Toyama, S. Murakami, J. Chem. Thermodyn. 18 (1986) 63-73.

[104] I. Nagata, K. Tamura, Fluid Phase Equilib. 24 (1985) 289–306.

[105] A.C. Galvao, A.Z. Francesconi, Thermochim. Acta 450 (2008) 81-86.

[106] F. Comelli, R. Francesconi, A. Bigi, K. Rubini, J. Chem. Eng. Data 51 (2006) 1711-1716.

[107] F. Kohler, J. Gaube, Pol. J. Chem. 54 (1980) 1987-1993.

[108] R. Fuchs, L. Krenzer, J. Gaube, Ber. Bunsenges. Phys. Chem. 88 (1984) 642-649.

[109] K. P. Shukla, A.A. Chialvo, J.M. Haile, Ind. Eng. Chem. Res. 27 (1988) 664-671.

[110] H. Wagner R.N. Lichtenthaler, Ber. Bunseges. Phys. Chem. 90 (1986) 69-70.

[111] M. Díaz Peña, B. Espino, R. Pérez, R.L. Arenosa, Boletín Soc. Quim. (Perú) 50 (1983) 64-68.

[112] A Saito, R. Tanaka, J. Chem. Thermodyn. 20 (1988) 859-865.

[113] J.A. González, I. García de la Fuente, J.C. Cobos, C. Casanova, H. V. Kehiaian, Thermochim. Acta 217 (1993) 57-69.

[114] I. García, J.C. Cobos, J.A. González, C. Casanova, M.J. Cocero, J. Chem. Eng. Data 33 (1988) 423-426.

[115] J.M. Pardo, C.A. Tovar, C.A. Cerdeiriña, E. Carballo, L. Romaní. J. Chem. Thermodyn. 31 (1999) 787-796.

[116] A. Tamir, A. Apelblat, M. Wagner, Fluid Phase Equilib. 6 (1981) 237-259.

[117] F. Becker, F. Hallauer, Int. DATA Ser., Sel. Data Mixtures, Ser. A 1 (1988) 43.

[118] L. Lepori, P. Gianni, E. Matteoli, J. Solution Chem. 42 (2013) 1263-1304.

[119]    F. Hevia, A. Cobos, J.A. González, I. García de la Fuente, V. Alonso, J. Solution Chem. 46 (2017) 150-174.

[120]    E. Tommila, E Lindell, M.L. Virtalaine, R. Laakso. Suomi Kemistil B. 42 (1969) 95-104.

[121]    A. Sacco, A. Kumar Rakshit, J. Chem. Thermodyn. 7 (1975) 257-261.

[122]    I. Cibulka, V.D. Nguyen, R.M. Holub. J. Chem. Thermodyn. 16 (1984) 159-164.

[123]    S. Vechi, M.S. Skaf, J. Chem. Phys. 123 (2005) 145507-8.

[124]    J.C.R. Reis, T.P. Iglesias, G. Douhéret, M.I. Davis, Phys. Chem. Chem. Phys. 11 (2009) 3977-3986.

[125]    J.A. González, L.F. Sanz, I. García de la Fuente, J.C. Cobos, J. Chem. Thermodyn. 91 (2015) 267-278.

[126]    F. Hevia, J.A. González, A. Cobos, I. García de la Fuente, L.F. Sanz. J. Chem. Thermodyn. 118 (2018) 175-187.

[127]    V. Alonso, J.A. González, I. García de la Fuente, J.C. Cobos, Thermochim. Acta, 551 (2013) 70-77.

[128]    C.F. Riadigos, R. Iglesias, M.A. Rivas, T.P. Iglesias, J. Chem. Thermodyn., 43 (2011) 275-283.

[129]    T.P. Iglesias, J.M. Forniés-Marquina, B. de Cominges, Mol. Phys. 103 (2005) 2639-2646.

[130]    J.A. Malecki, J. Chem. Phys. 43 (1965) 1351-1355.

[131]    C.M. Kinart, W.J. Kinart, Phys. Chem. Liq. 33 (1996) 151-158.

[132]    J. Lindberg, H.Hakalax, Finska Kemissamf Medd. 71 (1962) 97 (cf. [133])

[133]    C. Wohlfahrt, Static Dielectric Constants of Pure Liquids and Binary Liquid Mixtures. Landolt-Börnstein - Group IV Physical Chemistry Vol. 6. Springer Berlin Heidelberg, Berlin, 1991.

[134]    J. Guo-Zhu, Q. Jie, Fluid Phase Equilib. 365 (2014) 5-10

[135]    R. E. Powell, N. E. Roseveare, H. Eyring. Ind. Eng. Chem. 33 (1941) 430-435.

[136]    C. Moreau, G. Douhéret. Thermochim. Acta, 13 (1975) 385-392.

[137]    H. Eyring, M.S. Jones.  Significant Liquid Structure, Wiley, New York, 1969.

[138]    R.J. Martins, M.J.E. de M. Cardoso, O.E. Barcia. Ind. Eng. Chem. Res. 39 (2000) 849-854.

[139]    S. Chen,  Q. Lei, W. Fang. Fluid Phase Equilib., 234 (2005) 22-33.

[140]    P.S. Nikam, M.C. Jadhav, M. Hasan, J. Chem. Eng. Data 41 (1996) 1028-1031.

[141]    P.S. Nikam, L.N. Shirsat, M. Hasan J. Chem. Eng. Data 43 (1998) 732-737

[142]    C.-H. Tu, C.-Y. Liu, W.-F. Wang, Y.-T. Chou, J. Chem. Eng. Data 45 (2000) 450-456.

[143]    L.F. Sanz,  J.A. González, I. García de la Fuente, J.C. Cobos, Thermochim. Acta 631 (2016) 18-27.

[144] F. Mato, J. Berrueta, An. Quim. 74(1978) 1290–1293.

[145] L.F. Sanz, J.A. González, I. García de la Fuente, J.C. Cobos, J. Chem. Thermodyn. 80 (2015) 161-171.

[146] V. Majer, V. Svoboda, Enthalpies of Vaporization of Organic Compounds. Blackwell, Oxford, 1985.

[147] D. Dragoescu, M. Bendová, Z. Wagner, D. Gheorghe J. Mol. Liq. 223 (2016) 790-804

[148] K.F. Liu, W.T. Ziegler, J. Chem. Eng. Data 11 (1996) 187-189.

[149] W.V. Steele, R.D. Chirico, S.E. Knipmeyer, A. Nguyen, J. Chem. Eng. Data 42 (1997) 1021-1036.

[150] M. K. Patwari, R. K Bachu, S. Boodida, S. Nallani J. Chem. Eng. Data 54 (2009) 1069-1072.

[151] F.W. Evans, H.A. Skinner, Trans. Faraday Soc. 55 (1959) 255-259.

[152] Y. Uosaki, H. Matsumara, H. Ogiyama, T. Morisoshi, J. Chem. Thermodyn. 22 (1990) 797-801.

[153] N.D. Lebedeva, Y.A. Katin, G.Y. Akhmedova, Russ. J. Phys. Chem. (Engl. Transl.) 45 (1971) 1192-1193.

[154] S. Singh, V.K. Rattan, S. Kapoor, R. Kumar, A. Rampal, J. Chem. Eng. Data 50 (2005) 288-292

[155] A. Laínez, M. Rodrigo, A.H. Roux, J.-P. E. Grolier, E. Wilhelm, Calorim. Anal. Therm. 16 (1986) 153-158.

[156] I. Nagata, K. Tamura, Fluid Phase Equilib. 24 (1985) 289-306.

[157] I. Nagata, K. Tamura, J. Chem. Thermodyn. 20 (1988) 1101-1107.

[158] A.Z. Francesconi, D.H. Lanfredi Viola, J. Chem. Thermodyn. 47 (2012) 28-32.

[159] J.R. Khurma, O. Muthu, S. Munjai, B.D. Smith, J. Chem. Eng. Data 28 (1983) 119-123.

[160] M. Almasi, L. Mousavi, J. Mol. Liq. 163 (2011) 46-52.

[161] H. Feng, Y. Wang, J. Shi, G.C. Benson, B.C.-Y. Lu, J. Chem. Thermodyn. 23 (1991) 169-174.

[162] H. Iloukhani, H.A. Zarei, Phys. Chem. Liq. 46 (2008) 154-161.

[163] T.M. Letcher, P.K. Naicker, J. Chem. Thermodyn. 33 (2001) 1035-1047.

[164] I.L. Lee, C.H. Kang, B.-S. Lee, H.W. Lee, J. Chem. Soc. Faraday Trans. 86 (1990) 1477-1481.

[165] H. Sadek, R.M. Fuoss J. Am. Chem. Soc. 72 (1950) 301-306.

[166] R.S. Neyband, H. Zarei, J. Chem. Thermodyn. 80 (2015) 119-123.

[167] G. Dharmaraju, G. Narayanaswamy, G.K. Raman, J. Chem. Thermodyn. 12 (1980) 563-566.

[168] F. Mato, F. Fernández-Polanco, An. Quim. 71 (1975) 815.

[169] D. Decroocq, Bull. Soc. Chim. Fr. (1964) 127 (cf. [e4]).

[170] A. D'Aprano, I. D. Donato, J. Chem. Faraday Trans. I. 69 (1973) 1685.

[171] E. Tommila, E. Lindell, M.L. Virtalaine, R. Laakso, Suom Kemistil B42 (1969) 95 (cf. [133])

[172] A.N. Prajapati, A.D. Vyas, V.A. Rana, S.P. Bhatnagar, J. Mol. Liq. 151 (2010) 12-16.

[173] P.S. Nikam, M.C. Jadhav, M. Hasan, J. Chem. Eng. Data 40 (1995) 931-934.

[174] J. Liszi, Acta Chim, Acad. Sci. Hung. 84 (1975) 125 (cf. [133])

[175] M. Trampe, C.A. Eckert, J. Chem. Eng. Data 36 (1991) 112-118.

[176] F. Hevia, J.A. González, A. Cobos, I. García de la Fuente, C. Alonso-Tristán, Fluid Phase Equilib. 468 (2018) 18-28.

[177] B. Malesinska, Int. DATA Ser., Sel. Data Mixtures, Ser. A 3 (1974) 172-175.

[178] B. Malesinska, Int. DATA Ser., Sel. Data Mixtures, Ser. A 2 (1975) 75-78.

[179] I.A. McLure, A.T. Rodríguez, P.A. Ingham, J.F. Steele, Fluid Phase Equilib. 8 (1982) 271

[180] K. Tang, C. Zhou, X. An, W. Shen, J. Chem. Thermodyn. 31 (1999) 943-954.

[181] X. An, F. Jiang, H. Zhao, C. Chen, W. Shen, J. Chem. Thermodyn. 30 (1998) 751-760.

[182] T. Yin, A. Shi, J. Xie, M. Wang, Z. Chen, X. An, W. Shen, J. Chem. Eng. Data 59 (2014) 1312-1319

[183] X. An, P. Li, H. Zhao, W. Shen, J. Chem. Thermodyn. 30 (1998) 1049-1059.

[184] X. An, H. Zhao, F. Jiang, W. Shen, J. Chem. Thermodyn. 30 (1998) 21-26.

[185] X. An, H. Zhao, F. Jiang, C. Mao, W. Shen, J. Chem. Themodyn. 29 (1997) 1047-1054

[186] X. An, F. Jiu, H. Zhao, W. Shen, Acta Chim. Sin. 56 (1998) 141-146.

[187] J. Wang, X. An. N. Wang, H. Lv, S. Chai, W. Shen, J. Chem. Thermodyn. 40 (2008) 1638-1644.

[188] Y. Lei, Z. Chen. N. Wang, C. Mao, X. An, W. Shen, J. Chem. Thermodyn. 52 (2010) 864-872.

[189] C. Mao, N. Wang, X. Peng, X. An, W. Shen, J. Chem. Thermodyn. 40 (2008) 424-430.

[190] T. Yin, Y. Lei, M. Huang, Z. Chen, C. Mao, X. An, W. Shen, J. Chem. Thermodyn. 43 (2011) 656-663.

[191] Z. Chen, Y. Bai, T. Yin, X. An, W. Shen, J. Chem. Thermodyn. 54 (2012) 438-443.

[192] N. Wang, C. Mao, X. Peng, X. An, W. Shen, J. Chem. Thermodyn. 38 (2006) 732-738.

[193] V.P.Sazonov, V.V. Filippov, Izv. Vysh. Ucheb. Zaved Khim. Khim. Tekhnol. 18 (1975) 222 (cf. [58]).

[194] I.K. Zhuravleva, E.F. Zhuravlev, V.B. Mukhametshina, Z.Z. Khisametdinova, Zh. Fiz. Khim. 51 (1977) 1003 (cf. [58]).

[195] U. Domanska, M. Marciniak, J. Chem. Eng. Data 50 (2005) 2035-2044.

[196] V.P. Sazonov, L.V. Gudkina, Zh. Prikl. Khim. (Leningrad) 46 (1973) 1076 (cf. [58])

TABLE 1

Physical properties of pure compounds at 298.15 K[a]

| Compound | $V_m$ / cm$^3$·mol$^{-1}$ | $\mu$ /D | $\bar{\mu}$ | $\Delta\Delta H_{vap}$ | $\alpha_p$ / 10$^{-3}$·K$^{-1}$ | $\kappa_T$ / 10$^{-12}$·Pa$^{-1}$ |
|---|---|---|---|---|---|---|
| Ethanenitrile | 52.87 [1] | 3.53 [1] | 1.858 | 23.6 [146] | 1.41 [1] | 1070 [1] |
| Nitromethane | 53.96 [1] | 3.56 [1] | 1.855 | 28.6 [146] | 1.24 [1] | 738 [147,148] |
| Dimethylsulfoxide | 71.33 [1] | 4.06 [1] | 1.840 | 36.6 [1,146] | 0.928 [1] | 520 [1] |
| Sulfolane | 95.24 [121][b] | 4.81 [1] | 1.870 | 39.1 [146,149] | 0.8 [150] | 505 [1,150][c] |
| Benzonitrile | 103.06 [1] | 4.01 [1] | 1.512 | 17.7 [146,151] | 0.86 [1] | 611 [152] |
| Nitrobenzene | 102.74 [1] | 4.0 [1] | 1.510 | 18 [146,153] | 0.833 [1] | 424 [154,155] |

[a] $V_m$, molar volume; $\mu$, dipole moment; $\bar{\mu}$, effective dipole moment (equation 10); $\Delta\Delta H_{vap}$, differences between the standard enthalpy of vaporization at 298.15 for a compound with a given polar group and that of the corresponding homomorphic hydrocarbon (equation 9); $\alpha_p$, isobaric expansion coefficient; $\kappa_T$, isothermal compressibility factor; [b]value at 303.15 K; [c] using $\kappa_S$ (adiabatic compressibility) = 403·10$^{-12}$ Pa$^{-1}$ [150] and $C_p$ = 180 J·mol$^{-1}$·K$^{-1}$ [1]

TABLE 2

Excess molar functions at constant pressure, enthalpy, $H_m^E$, Gibbs energy, $G_m^E$, entropy, $TS_m^E$, and volume, $V_m^E$, and excess molar functions at constant volume, internal energy, $U_{Vm}^E$, and entropy, $TS_{Vm}^E$ for 1-alkanol(1) + polar compound(2) mixtures at equimolar composition and 298.15 K.

| 1-alkanol | $H_m^E$ / J·mol$^{-1}$ | $G_m^E$ / J·mol$^{-1}$ | $TS_m^E$ / J·mol$^{-1}$ | $V_m^E$ / cm$^3$·mol$^{-1}$ | $U_{Vm}^E$ / J·mol$^{-1}$ | $TS_{Vm}^E$ / J·mol$^{-1}$ |
|---|---|---|---|---|---|---|
| 1-alkanol + ethanenitrile | | | | | | |
| Methanol | 1086 [156] | 668[a] | 434 | −0.142 [122] | 1147 | 495 |
| Ethanol | 1502 [156] | 838[a] | 664 | −0.026 [122] | 1511 | 673 |
| 1-propanol | 1829 [97] | 956[a] | 873 | 0.055 [122] | 1810 | 854 |
| 1-butanol | 2044 [157] | 1038[a] | 1006 | 0.104 [122] | 2009 | 971 |
| 1-pentanol | 2005 [105] | | | 0.159 [122] | | |
| 1-hexanol | 2313 [105] | | | 0.205 [122] | | |
| 1-nonanol | 2671 [158] | | | | | |
| 1-decanol | 3032 [158] | | | | * | |
| 1-alkanol + nitromethane | | | | | | |
| Methanol | 1265 [96] | 1040 [159] | 225 | −0.152 [101] | 1323 | 283 |
| Ethanol | 1632 [96] | 1170 [159] | 462 | 0.026 [160] | 1623 | 453 |
| 1-propanol | 1911 [96] | 1254[b] | 758 | 0.235 [160] | 1826 | 672 |
| 1-butanol | 2131 [96] | 1338[a] | 793 | 0.333 [160] | 2011 | 673 |
| 1-hexanol | 2781[c] [161] | 1424[b,c] | 1357 | | | |
| 1-alkanol + DMSO | | | | | | |
| Methanol | −391 [98] | −718[a] | 327 | −0.588 [162] | −162 | 556 |
| Ethanol | 459 [98] | −391[a] | 850 | −0.228 [162] | 544 | 935 |
| 1-propanol | 762 [98] | −200[a] | 892 | −0.038 [162] | 776 | 906 |
| 1-butanol | 969 [81] | −45[a] | 1014 | 0.114 [162] | 927 | 972 |
| 1-pentanol | 1180 [81] | 62[a] | 1118 | | | |
| 1-hexanol | 1334 [81] | 129[a] | 1205 | | 1209 | 1080 |
| 1-octanol | 1502 [81] | | | | | |
| 1-decanol | 1606 [81] | | | | | |
| 1-alkanol + sulfolane | | | | | | |
| Methanol | 1551 [120][d] | 911[a,d] | 640 | −0.826 [121][d] | 1871 | 961 |
| Ethanol | 1971 [120][d] | 1128[a,d] | 843 | −0.697 [121][d] | 2223 | 1095 |

TABLE 2 (continued)

| | | | 1-alkanol + benzonitrile | | | |
|---|---|---|---|---|---|---|
| Methanol | 976 [163] | 862[a] | 114 | −0.358 [163] | 1105 | 243 |
| Ethanol | 1209 [163] | 948[a] | 261 | −0.330 [163] | 1325 | 377 |
| 1-propanol | 1454 [163] | 1196[a] | 258 | −0.262 [163] | 1545 | 349 |
| | | | 1-alkanol + nitrobenzene | | | |
| Methanol | 1109 [164] | 1256[a] | −147 | −0.386 [165] | 1273 | 17 |
| Ethanol | 1430 [166] | 1331[a] | 99 | | | |
| 1-propanol | 1807 [166] | 1380[a] | 427 | −0.192 [167][a] | 1883 | 503 |
| 1-butanol | 1946 [166] | 1401[a] | 545 | −0.117 [167][a] | 1992 | 591 |

[a]DISQUAC value; [b]extrapolated value from DISQUAC results; [c]value at 313.15 K; [d]value at 303.15 K

TABLE 3

Excess permittivies, $\varepsilon_r^E$, and Kirkwood correlations factors, $g_K$, at $\varphi_1 = 0.5$ and deviations of dynamic viscosities, $\Delta \eta$, at $x_1 = 0.5$ for 1-alkanol(1) + organic solvent(2) mixtures at temperature $T$.

| System | $T$/K | $\varepsilon_r^E$ | $g_K$ | $T$/K | $\Delta \eta$ /mPa·s |
|---|---|---|---|---|---|
| Methanol + ethanenitrile | 298.15 | 0.51 [37] | 1.38 | 298.15 | −0.08 [141] |
| ethanol + ethanenitrile | | | | 298.15 | −0.24 [141] |
| 1-propanol + ethanenitrile | 298.15 | −1.40 [168] | 1.13 | 298.15 | −0.54 [141] |
| 1-butanol + ethanenitrile | 298.15 | −1.48 [168] | 1.03 | 298.15 | −0.79 [141] |
| 1-pentanol + ethanenitrile | 298.15 | −1.67 [168] | | | |
| 1-hexanol + ethanenitrile | 298.15 | −1.57 [168] | | | |
| 1-heptanol + ethanenitrile | 298.15 | −1.48 [168] | 0.97 | | |
| Methanol + nitromethane | 293.15 | −1.74 [133,169][a] | 1.29 | 293.15 | −0.07 [142][a] |
| ethanol + nitromethane | 293.15 | −2.55 [133,169][a] | 1.13 | 293.15 | −0.24 [142][a] |
| 1-propanol + nitromethane | 293.15 | −3.20 [133,169][a] | 1.04 | 293.15 | −0.54 [142][a] |
| 1-butanol + nitromethane | | | | | |
| Methanol + DMSO | 298.15 | 4.04 [80] | 1.57 | 298.15 | −0.22 [140] |
| | | | | | −0.21 [80] |
| ethanol + DMSO | 298.15 | 0.066 [134] | 1.37 | 298.15 | −0.31 [140] |
| 1-Propanol + DMSO | 298.15 | 6.19 [131] | 1.55 | 298.15 | −0.40 [140] |
| | | −4.9 [134] | | | −0.52 [131] |
| | | −0.44 [132,133] | | | |
| 1-butanol + DMSO | 298.15 | −0.7 [132,133] | 1.24 | 298.15 | −0.51 [140] |
| Methanol + sulfolane | 298.15 | 0.602 [170] | 1.33 | 298.15 | −3.18 [170] |
| Ethanol + sulfolane | 293.15 | −0.85 [133,171][a] | 1.21 | | |
| Methanol + benzonitrile | 293.15 | 0.584 [133,169][a] | 1.37 | 303.15 | −0.031 [54][b] |
| ethanol + benzonitrile | 293.15 | 0.40 [133,169][a] | 1.21 | 303.15 | −0.140 [54][b] |
| 1-propanol + benzonitrile | 303.15 | −0.86 [172][b] | 1.13 | 303.15 | −0.315 [54][b] |
| 1-pentanol + benzonitrile | | | | 303.15 | −0.678 [54][b] |
| Methanol + nitrobenzene | 293.15 | −1.26 [133,169][a] | 1.32 | 298.15 | −0.055 [173] |
| ethanol + nitrobenzene | 293.15 | −2.46 [133,169][a] | 1.20 | 298.15 | −0.18 [173] |
| 1-propanol + nitrobenzene | 293.15 | −4.01 [133,169][a] | 1.10 | 298.15 | −0.40 [173] |
| 1-butanol + nitrobenzene | 293.15 | −4.23 [133,169][a] | | 298.15 | −0.56 [173] |
| 1-pentanol + nitrobenzene | 293.15 | −2.67 [133,174][a] | 1.13 | | |
| 1-heptanol + nitrobenzene | 293.15 | −0.90 [133,174][a] | 1.17 | | |

[a]value at 293.15 K; [b]value at 303.15 K

TABLE 4

First dispersive interchange coefficients, $C_{sh,1}^{DIS}$, for (s,h)[a] contacts in 1-alkanol + benzonitrile, or + nitrobenzene, or + sulfolane mixtures.

| $C_{sh,1}^{DIS}$ (contact) | Methanol | Ethanol | 1-propanol | 1-butanol |
|---|---|---|---|---|
| $C_{sh,1}^{DIS}$ (OH/CN) | 7 | 8.5 | 1[b] | |
| $C_{sh,1}^{DIS}$ (OH/NO$_2$) | 0.12[c] | 0.5 | 1 | 1 |
| $C_{sh,1}^{DIS}$ (OH/SO$_2$) | 3.5 | 3.5 | 3.3[d] | 3.3[d] |

[a] s = CN in benzonitrile; s = NO$_2$ in nitrobenzene; s = SO$_2$ in sulfolane; h = OH, in 1-alkanols); [b] [18] ; [c][17]; [d][13]

TABLE 5

Values of the Kirkwood-Buff integrals, $G_{ij}$, and of linear coefficients of preferential solvation, $\delta_{ij}$, at 298.15 for the mixtures 1-alkanol(1) + nitromethane(2) (NM), or + ethanenitrile(2) (EtN), or + nitrobenzene(2) (NTBz), or + benzonitrile(2) (BzCN), or + DMSO(2), or + sulfolane(2) (SULF) systems. Maximum/minimum $G_{ij}$ and $\delta_{ij}$ values are reported including the concentration ($x_1$) at which these maxima/minima are encountered. Otherwise, $G_{ij}$ and $\delta_{ij}$ results are determined at equimolar concentration.

| System | $G_{11}/$ cm$^3$ mol$^{-1}$ | $G_{22}/$ cm$^3$ mol$^{-1}$ | $G_{12}/$ cm$^3$ mol$^{-1}$ | $\delta_{21}/$ cm$^3$ mol$^{-1}$ | $\delta_{22}/$ cm$^3$ mol$^{-1}$ |
|---|---|---|---|---|---|
| Methanol + NM | 281 | 310 | −175 | −99 | 75 |
| | ($x_1 = 0.30$) | ($x_1 = 0.85$) | ($x_1 = 0.42$) | ($x_1 = 0.37$) | ($x_1 = 0.76$) |
| Ethanol + NM | 1222 | 1061 | −1192 | −590 | 555 |
| | ($x_1 = 0.43$) | ($x_1 = 0.48$) | ($x_1 = 0.46$) | ($x_1 = 0.45$) | ($x_1 = 0.47$) |
| 1-propanol + NM | 2319 | 1621 | −2000 | −996 | 847 |
| | ($x_1 = 0.36$) | ($x_1 = 0.39$) | ($x_1 = 0.37$) | ($x_1 = 0.37$) | ($x_1 = 0.38$) |
| 1-butanol + NM | 2941 | 3577 | −3244 | −1441 | 1619 |
| | ($x_1 = 0.37$) | ($x_1 = 0.41$) | ($x_1 = 0.39$) | ($x_1 = 0.38$) | ($x_1 = 0.40$) |
| Methanol + EtN | 21 | −24 | −83 | −26 | 15 |
| 1-propanol + EtN | −9 | 110 | −163 | −38 | 68 |
| 1-butanol + EtN | −14 | 246.5 | −221 | −51.7 | 117 |
| 1-hexanol + EtN | −45 | 561 | −309 | −66 | 217 |
| 1-octanol + EtN | 1264 | 1266 | −1170 | −378 | 486 |
| | ($x_1 = 0.14$) | ($x_1 = 0.35$) | ($x_1 = 0.23$) | ($x_1 = 0.23$) | ($x_1 = 0.33$) |
| Methanol + NTBZ | 480 | 1576 | −585 | −207 | 205 |
| | ($x_1 = 0.50$) | ($x_1 = 0.93$) | ($x_1 = 0.86$) | ($x_1 = 0.29$) | ($x_1 = 0.86$) |
| 1-propanol + NTBZ | 1610 | 4046 | −2487 | −897 | 1339 |
| | ($x_1 = 0.63$) | ($x_1 = 0.92$) | ($x_1 = 0.68$) | ($x_1 = 0.65$) | ($x_1 = 0.69$) |
| 1-butanol + NTBZ | 2650 | 5501 | −3758 | −1457 | 2125 |
| | ($x_1 = 0.58$) | ($x_1 = 0.64$) | ($x_1 = 0.62$) | ($x_1 = 0.62$) | ($x_1 = 0.63$) |
| Methanol + BZCN | 275 | 308 | −349 | −125 | 101 |
| | ($x_1 = 0.66$) | ($x_1 = 0.86$) | ($x_1 = 0.76$) | ($x_1 = 0.67$) | ($x_1 = 0.79$) |

TABLE 5 (continued)

| | | | | | |
|---|---|---|---|---|---|
| Ethanol + BZCN | 326 | 340 | −402 | −164 | 146 |
| | ($x_1 = 0.55$) | ($x_1 = 0.78$) | ($x_1 = 0.67$) | ($x_1 = 0.59$) | ($x_1 = 0.68$) |
| 1-propanol + BZCN | 401 | 1040 | −540 | −205 | 255 |
| | | ($x_1 = 0.87$) | ($x_1 = 0.70$) | | ($x_1 = 0.73$) |
| Methanol + DMSO | −52 | −91 | −30 | 5 | −15 |
| Ethanol + DMSO | −70 | −89 | −45 | 6 | −11 |
| 1-Propanol + DMSO | −87 | −80 | −59 | 7 | −5 |
| 1-butanol + DMSO | −104 | −66 | −73 | 8 | 2 |
| Methanol + sulfolane | 236.4 | −65.8 | −155 | −97.9 | 22.4 |
| Ethanol + sulfolane | 615.9 | 730 | −589 | −303.7 | 303.6 |
| | ($x_1 = 0.56$) | ($x_1 = 0.72$) | ($x_1 = 0.65$) | ($x_1 = 0.59$) | ($x_1 = 0.66$) |
| 1-propanol + sulfolane | 1088 | 1267 | −1234 | −561 | 589 |
| | ($x_1 = 0.53$) | ($x_1 = 0.62$) | ($x_1 = 0.58$) | ($x_1 = 0.55$) | ($x_1 = 0.59$) |
| 1-butanol + sulfolane | 2594 | 2812 | −2786 | −1332 | 1382 |
| | ($x_1 = 0.50$) | ($x_1 = 0.54$) | ($x_1 = 0.52$) | ($x_1 = 0.51$) | ($x_1 = 0.52$) |

TABLE 6

Partial molar excess enthalpies,[a] $H_1^{E,\infty}$, at $T = 298.15$ K at atmospheric pressure for solute(1) + organic solvent(2) mixtures, and enthalpies of 1-alkanol/solvent (polar group X) interactions, $\Delta H_{OH-CN}$, for 1-alkanol(1) + organic solvent(2) mixtures.

| System | $H_1^{E,\infty}$ /kJ·mol$^{-1}$ | $\Delta H_{OH-X}$ /kJ·mol$^{-1}$ |
|---|---|---|
| Ethanenitrile(1) + C$_6$H$_{12}$(2) | 15.0 [175] | |
| DMSO(1) + alkane(2) | 25.0 [94] | |
| Sulfolane(1) + hexane(2) | 32.0[a] [95] | |
| 1-pentanol(1) + ethanenitrile(2) | 11.9 [105] | −26.3 |
| 1-hexanol(1) + ethanenitrile(2) | 13.9 [105] | −24.3 |
| Methanol(1) + DMSO(2) | −1.58 [98] | −49.8 |
| ethanol(1) + DMSO(2) | 0.90 [98] | −47.3 |
| 1-propanol(1) + DMSO(2) | 2.51 [81] | −45.7 |
| 1-butanol(1) + DMSO(2) | 3.34 [81] | −44.9 |
| 1-pentanol(1) + DMSO(2) | 5.33 [81] | −42.9 |
| 1-hexanol(1) + DMSO(2) | 6.55 [81] | −41.7 |
| 1-heptanol(1) + DMSO(2) | 7.87 [81] | −40.3 |
| 1-octanol(1) + DMSO(2) | 9.55 [81] | −38.7 |
| 1-decanol(1) + DMSO(2) | 12.0 [81] | −36.2 |
| Methanol(1) + sulfolane(2) | 5.4[b] [15] | −49.8 |
| ethanol(1) + sulfolane(2) | 6.2[b] [15] | −49.0 |

[a]from activity coefficients at infinite dilution for the sulfolane + hexane system in the temperature range (334.6-341.4) K. [b]value at t 303.15 K

TABLE 7

Results for the regressions of the type $H_m^E(x_1=0.5; T/K=298.15) = m + nY$ [a] for 1-alkanol(1) + organic solvent(2) mixtures.

| Solvent | $n_{OH}$ [b] | Regression | $r$ [c] | $\sigma$ /J·mol$^{-1}$ [d] |
|---|---|---|---|---|
| Ethanenitrile | 1,2,3,4,6,9,10 | $H_m^E = 5790 + 0.140\,\Delta H_{OH\text{-}CN}$ | 0.983 | 134 |
|  | 1,2,3,4 | $H_m^E = 8065 + 0.215\,\Delta H_{OH\text{-}CN}$ | 0.990 | 71 |
| Nitromethane | 1,2,3,4 | $H_m^E = 5286 + 0.132\,\Delta H_{OH\text{-}NO2}$ | 0.995 | 44 |
| Dimethyl sulfoxide | 1,2,3,4 | $H_m^E = 13339 + 0.275\,\Delta H_{OH\text{-}SO}$ | 0.992 | 93 |
|  | 5,6,7,8,10 | $H_m^E = 3819 + 0.061\,\Delta H_{OH\text{-}SO}$ | 0.980 | 46 |
| Benzonitrile | 1,2,3 | $H_m^E = 5909 + 0.183\,\Delta H_{OH\text{-}CN}$ | 0.976 | 74 |
| Nitrobenzene | 1,2,3,4 | $H_m^E = 4763 + 0.156\,\Delta H_{OH\text{-}NO2}$ | 0.996 | 44 |
| Ethanenitrile | 1,3,4,6 | $H_m^E = 512 - 28.75\,\delta_{21}$ | 0.943 | 214 |
| Nitromethane | 1,2,3,4 | $H_m^E = 1227 - 0.65\,\delta_{21}$ | 0.996 | 42 |
| Dimethyl sulfoxide | 1,2,3,4 | $H_m^E = -2339 + 438.3\,\delta_{21}$ | 0.946 | 238 |
| Benzonitrile | 1,2,3 | $H_m^E = 229 + 5.97\,\delta_{21}$ | 1 | 0.02 |

[a] $Y = \Delta H_{OH\text{-}X}$/J·mol$^{-1}$ (enthalpies of the 1-alkanol/solvent (polar group X) interactions; values from Table 6 and [16,17]); $m$ units: J·mol$^{-1}$; $Y = \delta_{21}$/cm$^3$·mol$^{-1}$ (linear coefficients of preferential solvation for molecules of polar compound around a central 1-alkanol molecule; values from Table 5); $m$ units: J·mol$^{-1}$; $n$ units: J·cm$^{-3}$; [b] number of C atoms in the 1-alkanol; [c] coefficient of linear regression; [d] standard deviation, $\sigma_r(H_m^E) = \{\dfrac{1}{N-2}\sum(H_{m,exp}^E - H_{m,calc}^E)^2\}^{1/2}$

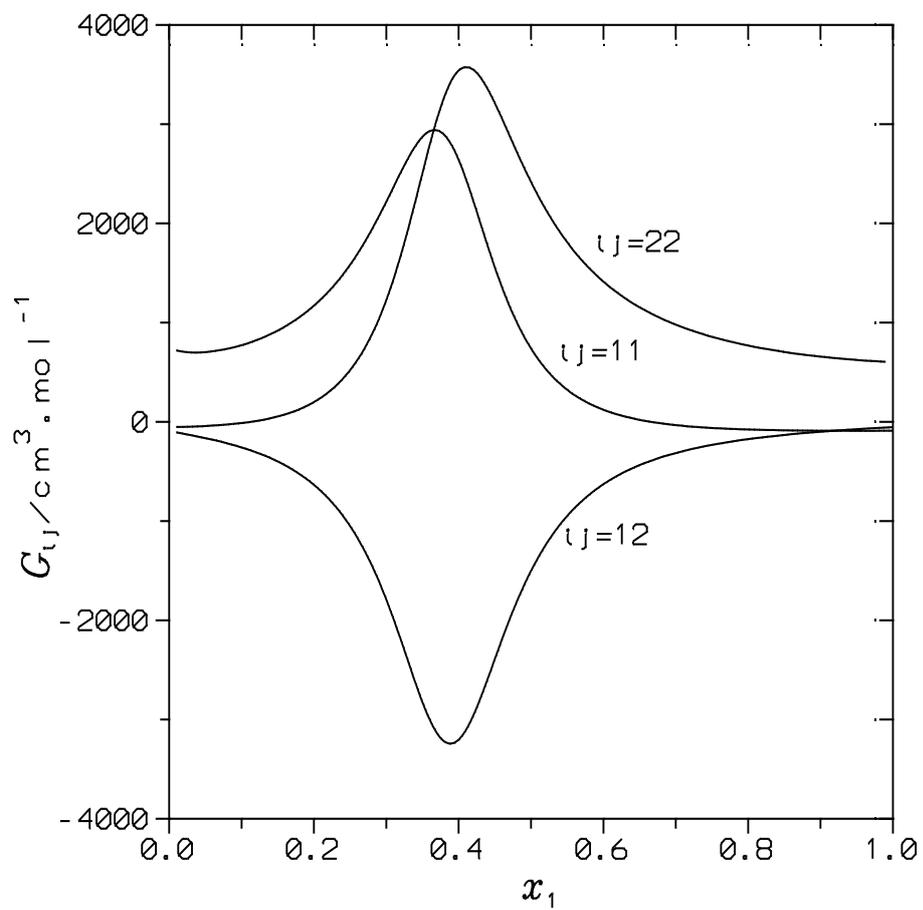

**Figure 1**  Kirkwood-Buff integrals, $G_{ij}$, for the 1-butanol(1) + nitromethane(2) system at 298.15 K

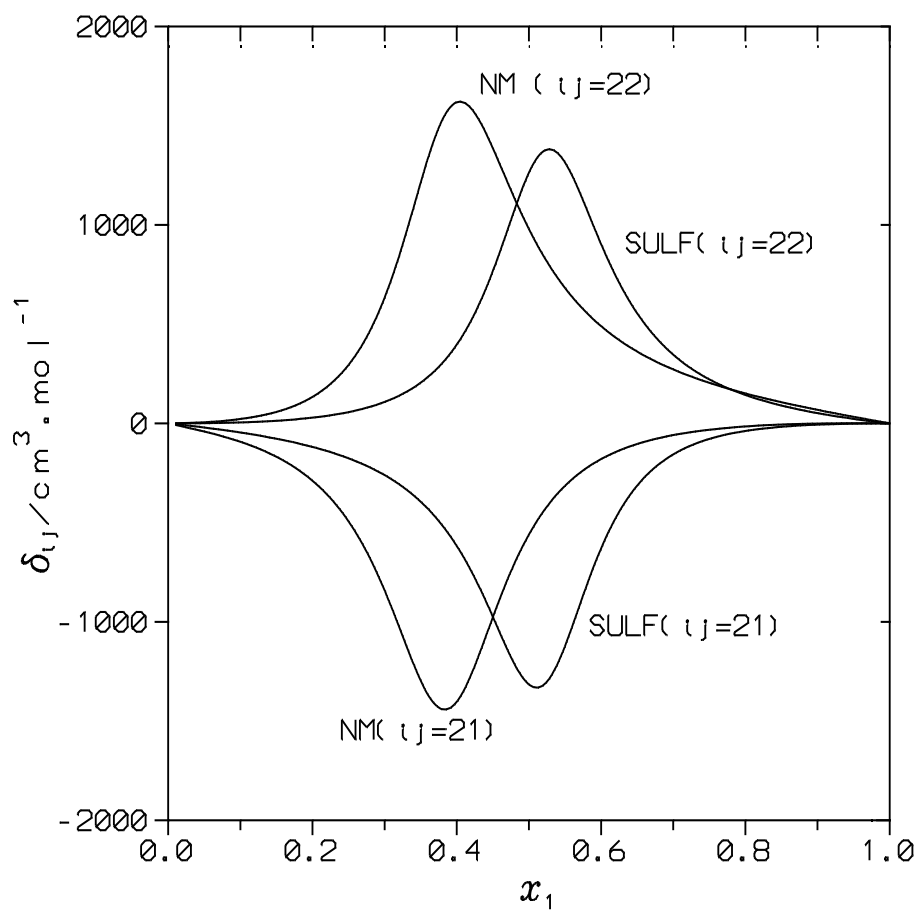

**Figure 2**    Linear coefficients of preferential solvation, $\delta_{ij}$, for 1-butanol(1) + nitromethane(2) (NM), or + sulfolane(2) (SULF) systems at 298.15 K

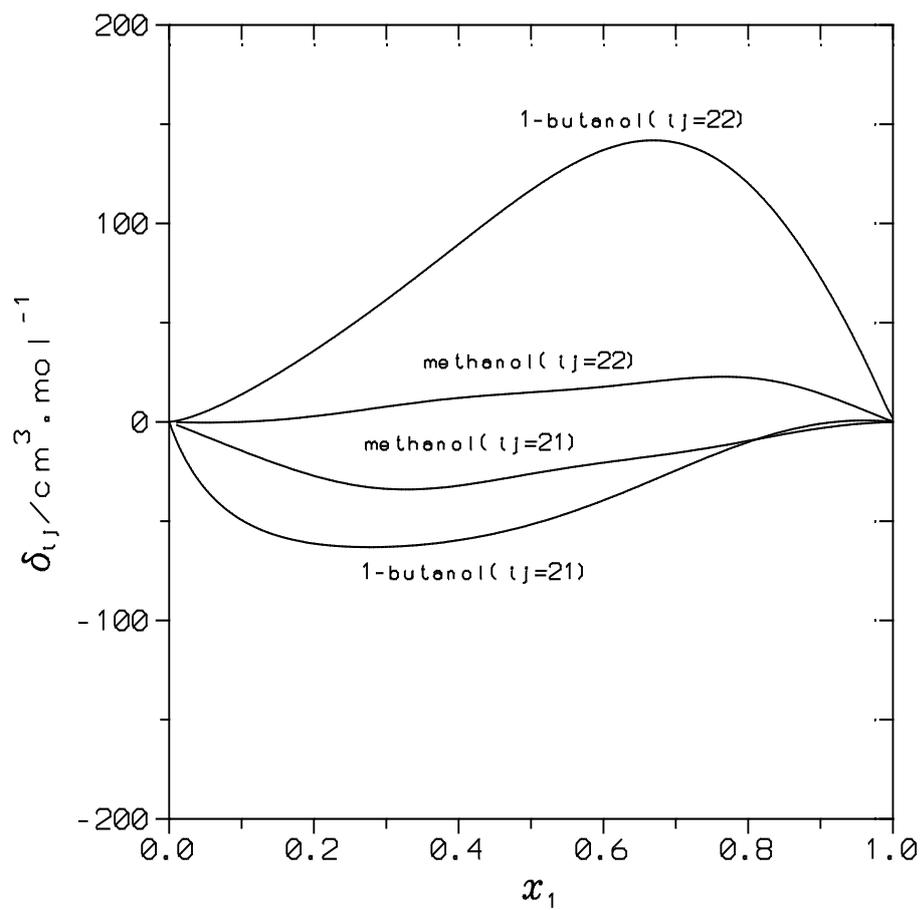

**Figure 3**  Linear coefficients of preferential solvation, $\delta_{ij}$, for 1-alkanol(1) + ethanenitrile(2) systems at 298.15 K.

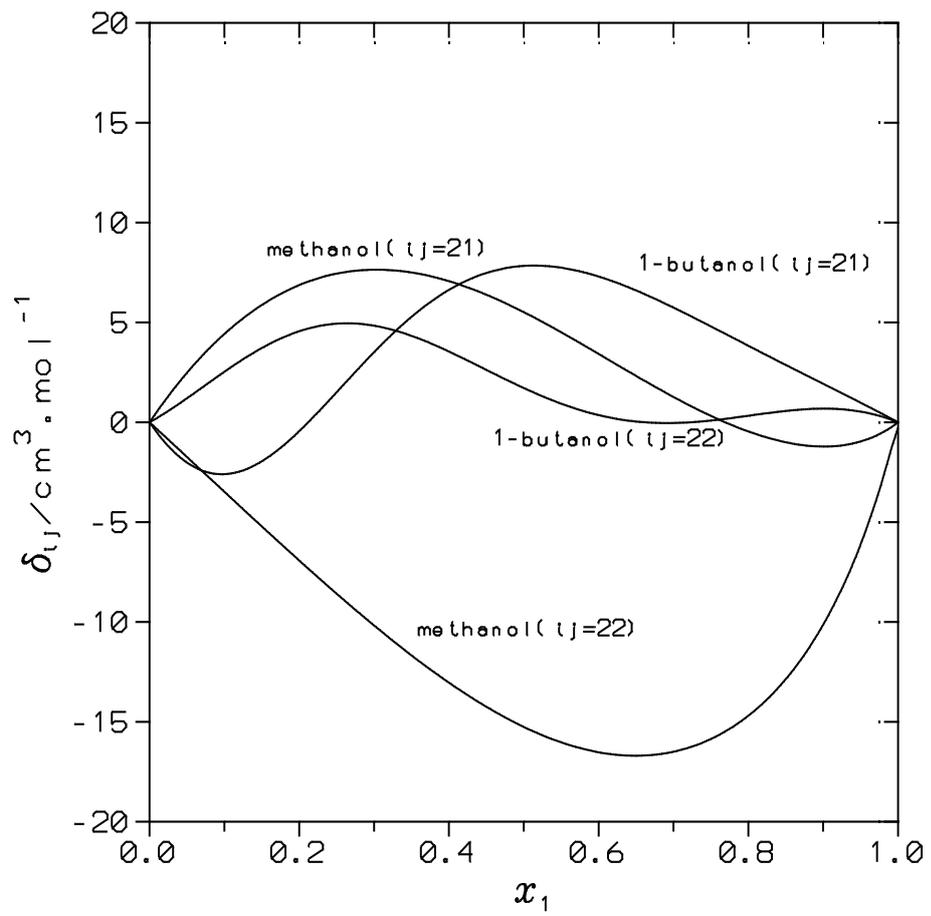

**Figure 4**  Linear coefficients of preferential solvation, $\delta_{ij}$, for 1-alkanol(1) + DMSO(2) systems at 298.15 K

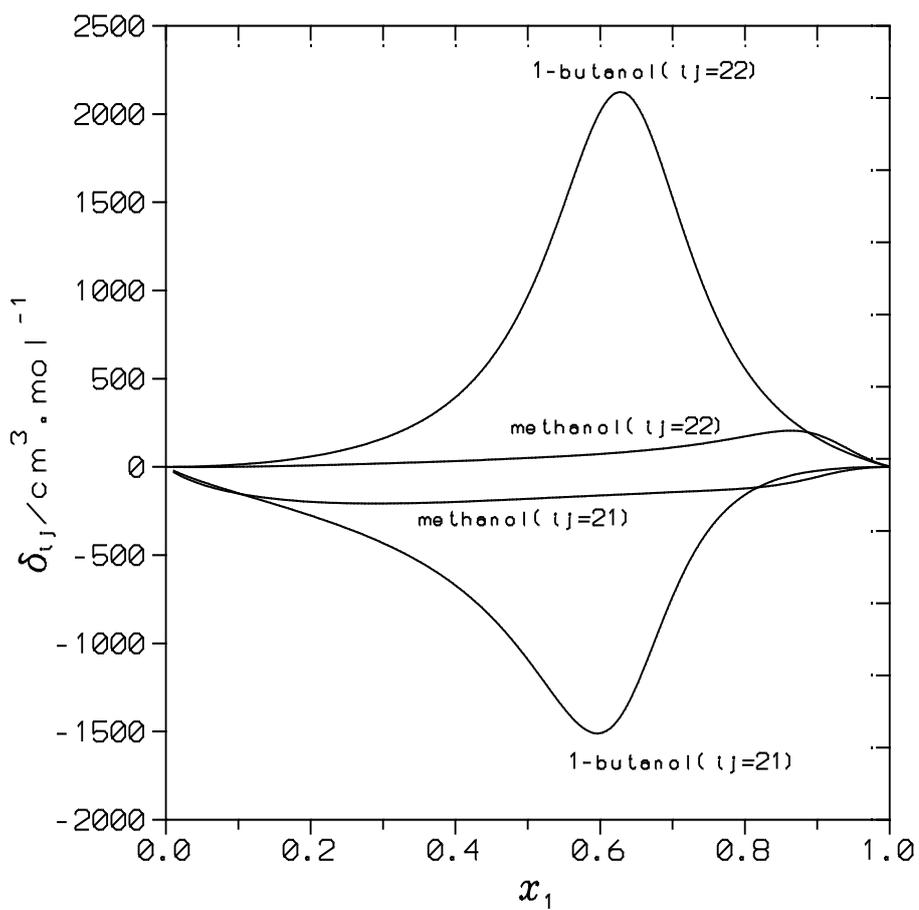

**Figure 5** Linear coefficients of preferential solvation, $\delta_{ij}$, for 1-alkanol(1) + nitrobenzene(2) systems at 298.15 K

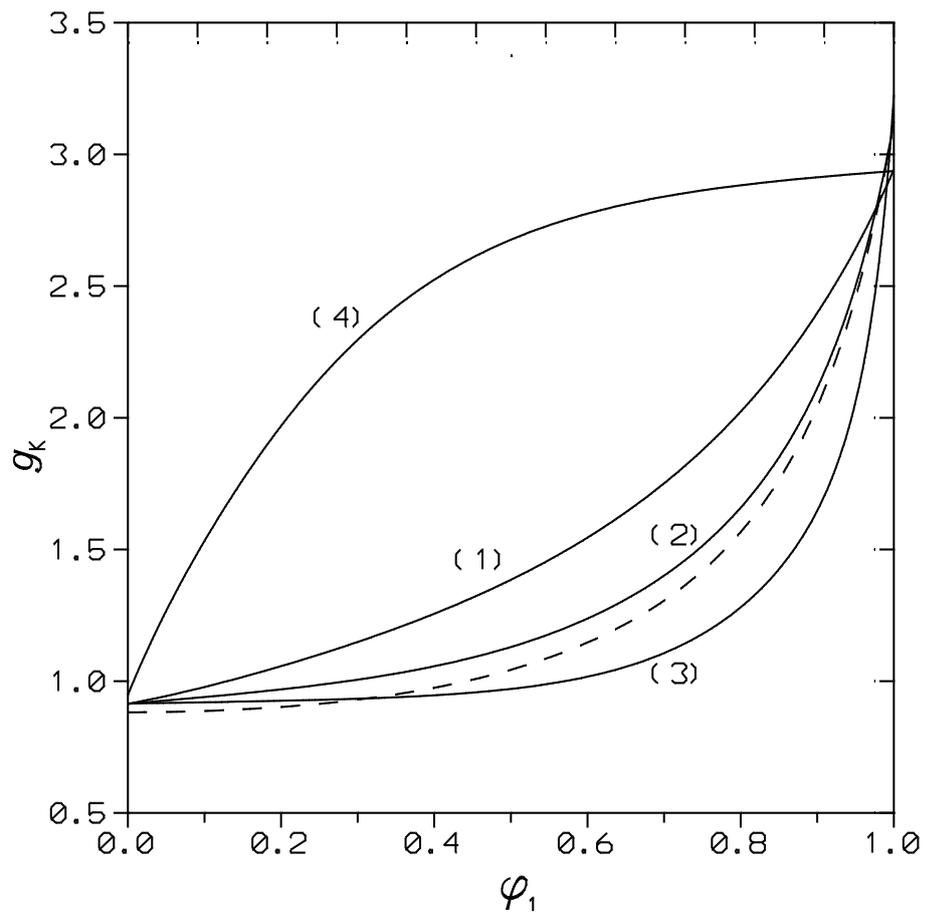

**Figure 6**   Kirkwood correlation factors, $g_K$, for the systems at 298.15 K: (1), methanol(1) + ethanenitrile(2); (2), 1-propanol(1) + ethanenitrile(2); (3), 1-heptanol(1) + ethanenitrile(2); (4), methanol(1) + hexylamine(2); (----) [176], 1-propanol(1) + nitromethane(2) ($T$ = 293.15 K).

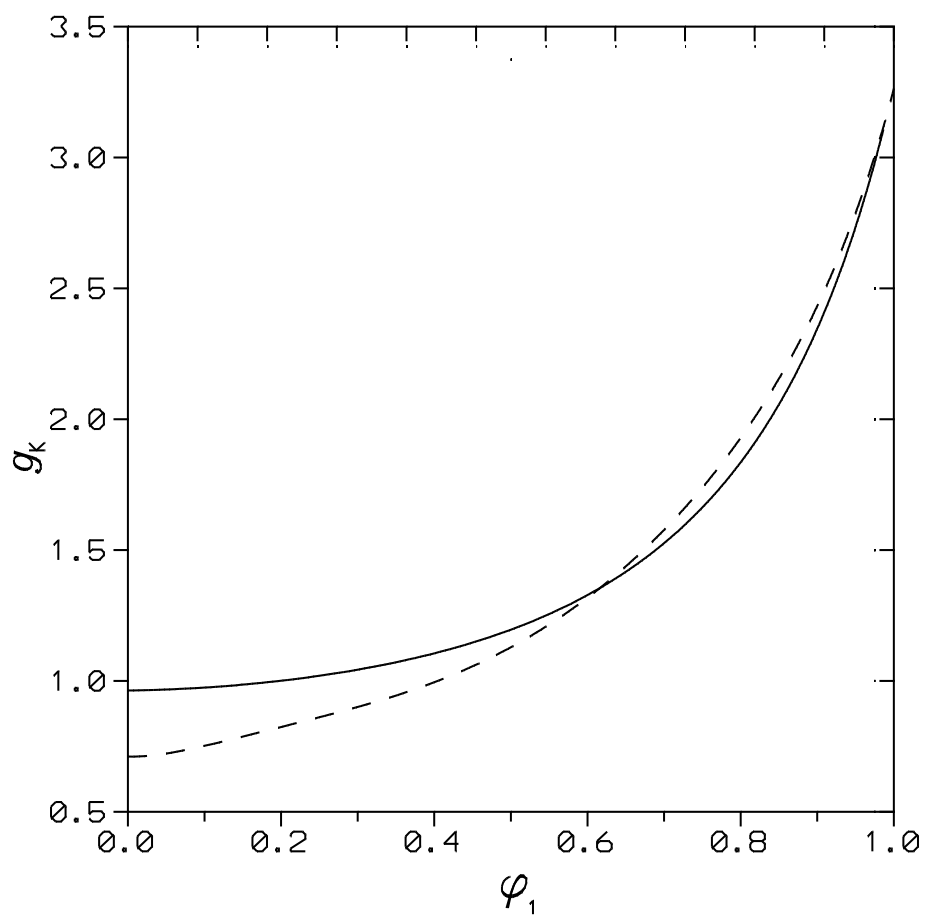

**Figure 7**  Kirkwood correlation factors, $g_K$, for the systems 1-propanol(1) + nitrobenzene(2) (solid line, $T = 293.15$ K), or + benzonitrile(2) (dashed line, $T = 303.15$ K).

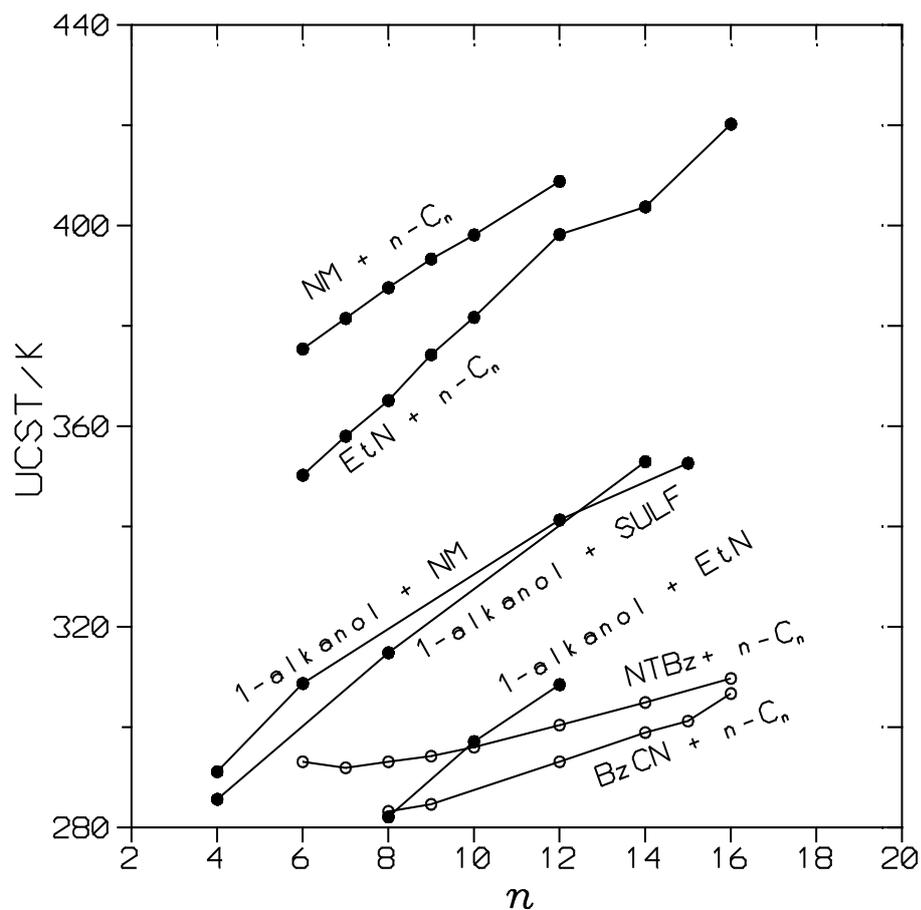

**Figure 8**  Upper critical solution temperatures (UCST) vs. $n$, the number of C atoms in $n$-alkanes or in 1-alkanols for mixtures investigated in this work. Lines are only for the aid of the eye. Nitromethane (NM) [59,177,178], or ethanenitrile (EtN) [179], or nitrobenzene (NTBz) [180-186], or + benzonitrile (BzCN) [187-192] + $n$-alkane; 1-alkanol + NM [58,193,194], or + EtN [58, 195,196], or + sulfolane (SULF) [13].

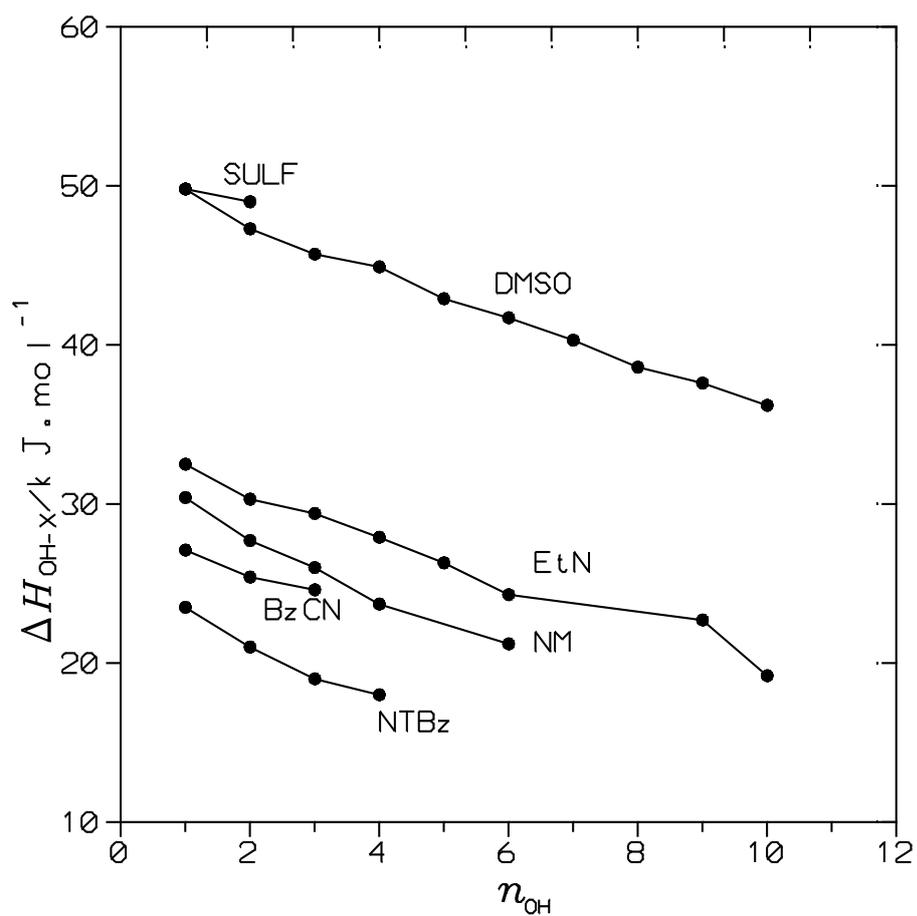

**Figure 9**  Values of the enthalpy of the interactions, $\Delta H_{OH-X}$, between 1-alkanols and a polar compound including a functional group X (= c-$SO_2$; $NO_2$; CN; SO) (Table 6, [16,17]) vs. $n_{OH}$ the number of C atoms in the 1-alkanol. Lines are only for the aid of the eye.

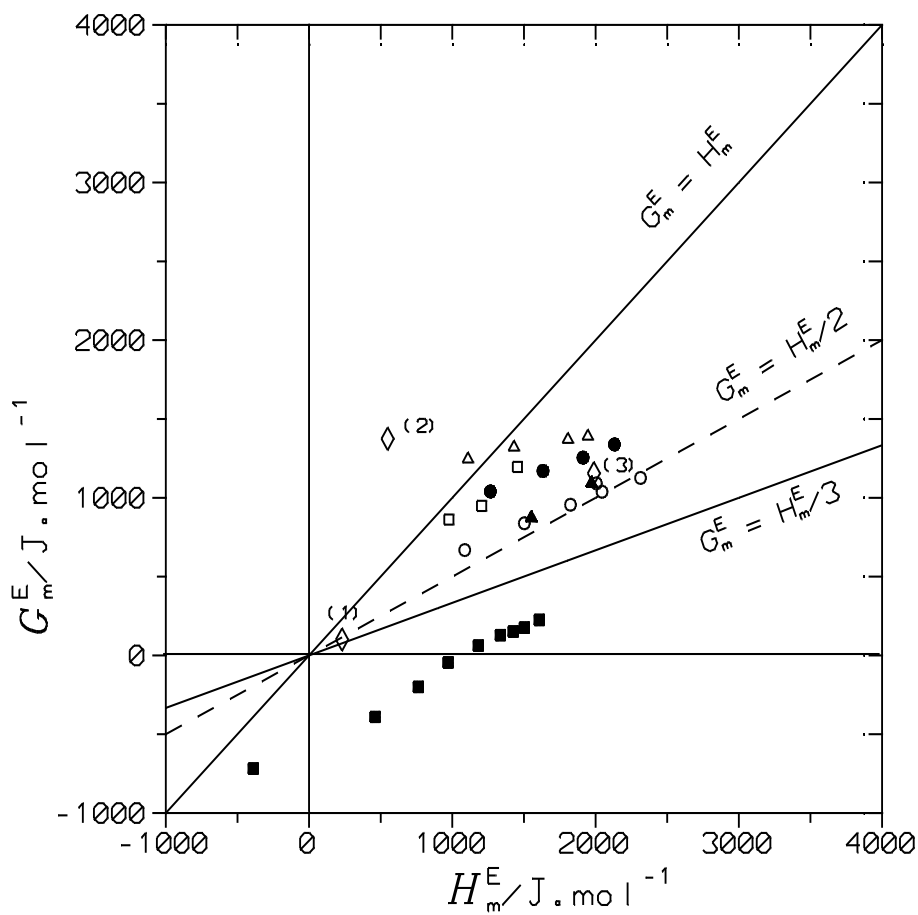

**Figure 10**  $G_m^E$ vs $H_m^E$ diagram for mixtures investigated in this work. Values of the excess molar functions are given at equimolar composition and 298.15 K (see Table 2). Symbols: (●), 1-alkanol + NM; (O), 1-alkanol + EtN; (Δ), 1-alkanol + NTBz; (□), 1-alkanol + BzCN; (▲), 1-alkanol + sulfolane ($T$ = 303.15 K); (■),1-alkanol + DMSO. For comparison are also represented the following mixtures (◇): (1) cyclohexane + hexane [110,111]; (2) ethanol + hexane [92,100]; (3) dimethyl carbonate + heptane [113,114].

SUPPLEMENTARY MATERIAL

# CHARACTERIZATION OF 1-ALKANOL + STRONGLY POLAR COMPOUND MIXTURES FROM THERMOPHYSICAL DATA AND THE APPLICATION OF THE KIRKWOOD-BUFF INTEGRALS AND KIRKWOOD-FRÖHLICH FORMALISMS.

JUAN ANTONIO GONZÁLEZ*, FERNANDO HEVIA, LUIS FELIPE SANZ, ISAÍAS GARCÍA DE LA FUENTE AND JOSÉ CARLOS COBOS

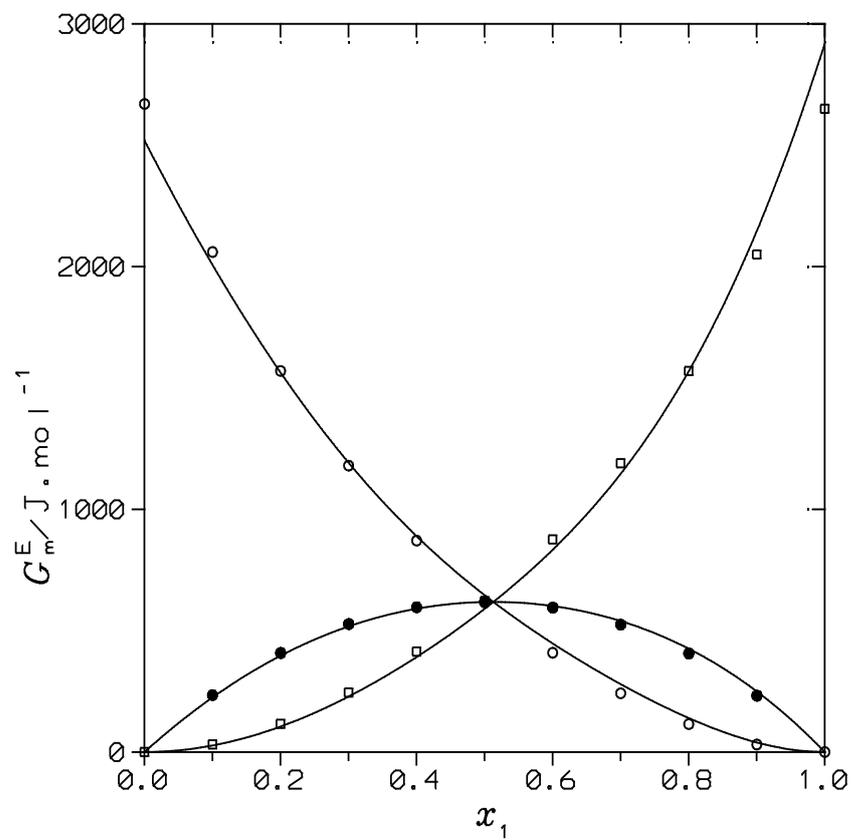

FIGURE S1    $G_m^E$ and $\mu_i^E$ for the methanol(1) + ethanenitrile(2) system at 328.15 K. Symbols, experimental values [1]: (●), $G_m^E$; (O), $\mu_1^E$, (□), $\mu_2^E$. Solid lines, DISQUAC calculations

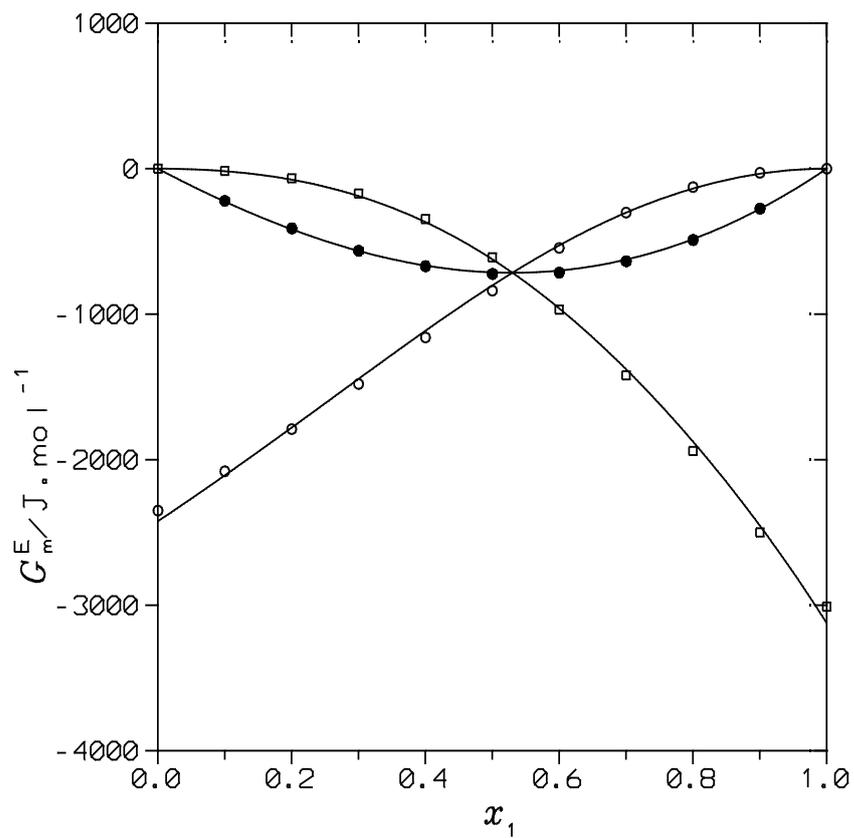

FIGURE S2 $G_m^E$ and $\mu_i^E$ for the methanol(1) + dimethyl sulfoxide(2) system at 293.15 K. Symbols, experimental values [2]: (●), $G_m^E$; (O), $\mu_1^E$, (□), $\mu_2^E$. Solid lines, DISQUAC calculations

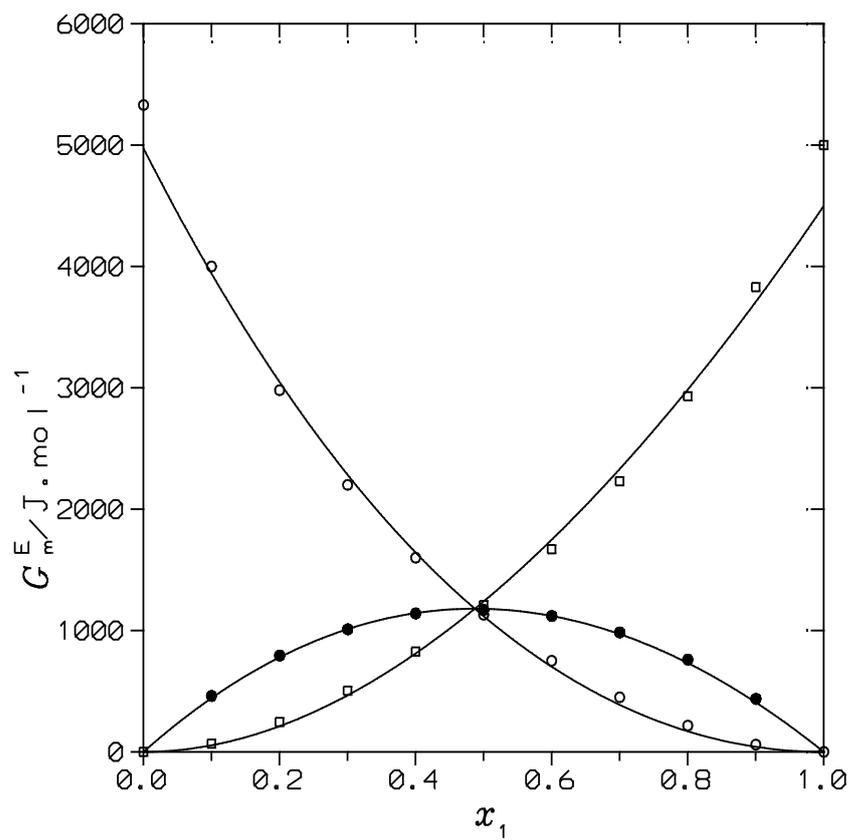

FIGURE S3  $G_m^E$ and $\mu_i^E$ for the ethanol(1) + nitromethane(2) system at 298.15 K. Symbols, experimental values [3]: (●), $G_m^E$; (O), $\mu_1^E$, (□), $\mu_2^E$. Solid lines, DISQUAC calculations

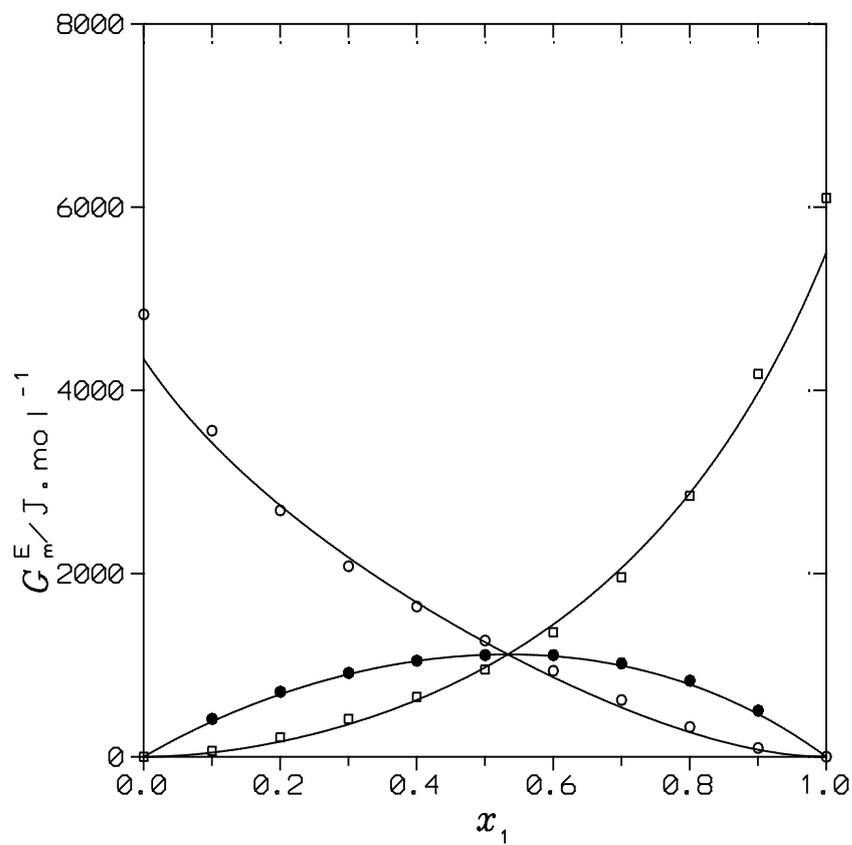

FIGURE S4   $G_m^E$ and $\mu_i^E$ for the ethanol(1) + sulfolane(2) system at 308.15 K. Symbols, experimental values [4]: (●), $G_m^E$; (O), $\mu_1^E$, (□), $\mu_2^E$. Solid lines, DISQUAC calculations

## References for supplementary material

[1]   I. Nagata, K. Katoh, J. Koyabu, Thermochim. Acta, 47 (1981) 225-233.

[2]   K. Quitzsch, H. Ulbrecht, G. Geiseler, Z. Phys. Chem. (Leipzig) 234 (1967) 33-43.

[3]   J.R. Khurma, O. Muthu, S. Munjai, B.D. Smith, J. Chem. Eng. Data 28 (1983) 119-123.

[4]   E. Tommila, E. Lindell, M.L. Virtalaine, R.Laakso, Suom Kemistil 42 (1969) 95-104.